\documentclass[letterpaper]{article}
\usepackage{geometry}
\usepackage{setspace}
\usepackage[articletitle=true]{achemso}
\usepackage{graphicx}
\usepackage{float}
\usepackage{caption}

\newfloat{scheme}{htbp}{los}
\floatname{scheme}{Scheme}

\newfloat{graph}{htbp}{loh}
\floatname{graph}{Graph}

\newfloat{chart}{htbp}{loc}
\floatname{chart}{Chart}

\usepackage{amsmath,amssymb,bm,mathtools}
\usepackage[version=4]{mhchem}
\usepackage{siunitx}
\usepackage{booktabs}
\usepackage{array}
\usepackage{multirow}

\usepackage[hidelinks]{hyperref}

\usepackage{authblk}

\title{Scalable on-chip integration of diamond color centers \\for cryogenic quantum photonics}
\usepackage{color}
\usepackage{graphicx}
\graphicspath{{C:/Users/scott/OneDrive/document/kosakaLab/paper/Chip_integrated_diamond_cc/figures/}}

\usepackage{amsmath,amssymb,amsfonts}
\usepackage{bm}            % 太字ベクトル
\usepackage{siunitx}       % 単位 ( \SI{1.0}{\kelvin} など )
\usepackage{braket}        % Dirac 記法
\usepackage{physics}       % \dv, \pdv など（不要なら削除）
\usepackage{graphicx}      % 画像
\usepackage{microtype}     % 微調整
\usepackage[dvipsnames]{xcolor}
\usepackage{hyperref}      % クリッカブルな参照
\hypersetup{
	colorlinks=true,
	linkcolor=MidnightBlue,
	citecolor=MidnightBlue,
	urlcolor=MidnightBlue
}

\author[1]{H.~Kurokawa\thanks{Email: kurokawa-hodaka-hm@ynu.ac.jp}}
\author[2]{K.~Sato}
\author[2]{M.~Kamata}
\author[3]{S.~Ishida}
\author[4]{H.~Matsukiyo}
\author[4]{N.~Pholsen}
\author[4]{M.~Nishioka}
\author[4,5]{S.~Ji}
\author[4]{H.~Otsuki}
\author[2]{S.~Hachuda}
\author[2]{M.~Kunii}
\author[2]{T.~Tamanuki}
\author[6]{K.~Kimura}
\author[6]{K.~Takenaka}
\author[1]{Y.~Sekiguchi}
\author[1,6]{S.~Onoda}
\author[1,3,4]{S.~Iwamoto}
\author[1,2]{T.~Baba}
\author[1,2]{H.~Kosaka\thanks{Email: kosaka-hideo-yp@ynu.ac.jp}}

\affil[1]{Quantum Information Research Center, Institute of Advanced Sciences, Yokohama National University, 79-5 Tokiwadai, Hodogaya, Yokohama, Kanagawa, 240-8501, Japan}

\affil[2]{Graduate School of Engineering Science, Yokohama National University, 79-5 Tokiwadai, Hodogaya, Yokohama, Kanagawa, 240-8501, Japan}

\affil[3]{Research Center for Advanced Science and Technology (RCAST), The University of Tokyo, 4-6-1 Komaba, Meguro-ku, Tokyo, 153-8904, Japan}

\affil[4]{Institute of Industrial Science (IIS), The University of Tokyo, 4-6-1 Komaba, Meguro-ku, Tokyo, 153-8904, Japan}

\affil[5]{Research Institute for Semiconductor Engineering, Hiroshima University, 1-4-2 Kagamiyama, Higashi-Hiroshima, Hiroshima, 739-8527, Japan}

\affil[6]{Quantum Materials and Applications Research Center (QUARC), QST, 1233 Watanuki-machi, Takasaki, Gunma, 370-1292, Japan}

\begin{document}
	\maketitle
	
	\begin{abstract}
	Chip integration of quantum emitters is a crucial milestone for scalable quantum photonic information processing. Among optically active defect centers for quantum photonics, diamond color centers are promising because of their long spin coherence times and high photon emission rates. However, for a coherent-photon emission, they typically require a cryogenic environment to protect optical coherence from thermal phonons, which makes chip integration challenging. In this paper, we develop a chip-integrated diamond photonic crystal cavity embedding an ensemble of nitrogen-vacancy (NV) centers. We confirm cryogenic operation by observing Purcell enhancement of NV-center emission via an edge-coupled optical fiber. This result demonstrates successful integration of diamond color centers, a photonic crystal cavity, and an optical waveguide-fiber package, representing a key step toward scalable diamond-based quantum communication platforms.
	\end{abstract}
	
\section*{Keywords}
Diamond, Nitrogen vacancy, Integrated photonics

	%%%%%%%%%%%%%%%%%%%%
	\section{}
	With advances in optically active quantum defects and improvements in nanofabrication technology, the on-chip integration of these defects has attracted increasing attention for scalable quantum photonic technologies, including single-photon sources  \cite{Chanana2022,Prabhu2023,Larocque2024}, quantum memories \cite{Mouradian2015,Wan2020, Li2024c}, quantum frequency transducers \cite{Kurokawa2022,Kim2023b,Xie2024,Kurokawa2025}, and quantum computers \cite{Lukin2020,Oberg2025}. Furthermore, when combined with photonic integrated circuits using heterogeneous integration techniques \cite{Harris2016,Davanco2017,Bogdanov2017,Lenzini2018, Wang2020d,Elshaari2020,  Pelucchi2022a, Labonte2024}, on-chip quantum signal processing on hybrid classical-quantum photonic platforms becomes feasible, extending both the capability and scalability of quantum information processing.

	Among various quantum defects and host materials for quantum photonics \cite{Harris2016,Lenzini2018	,Lukin2020,Labonte2024}, color centers in diamond have been extensively investigated owing to their long spin coherence times \cite{Balasubramanian2009} and high photon emission rates \cite{Ruf2021}. Particularly, nitrogen-vacancy (NV) centers \cite{Manson2005,Maze2011} attract broad interest owing to their potential applications in both quantum sensing \cite{Degen2017} and quantum communication \cite{Bernien2013a}. For quantum communication based on coherent single photons emitted from an NV center, cryogenic temperatures below 10 K are required to suppress incoherent thermal phonon occupation \cite{Fu2009a}. Additionally, the integration of color centers into nanophotonic structures has been widely adopted to enhance photon emission and collection efficiency \cite{Li2015c,Sipahigil2016, Parker2024}. Therefore, the operation of integrated nanophotonic structures at cryogenic temperatures is an essential requirement for practical diamond integrated devices. Similar requirements apply to other color-center systems, such as group-IV vacancy centers in diamond \cite{Sipahigil2016, Parker2024},and defects in silicon carbide \cite{Crook2020}, and silicon \cite{Redjem2023,Johnston2024}.

	This study develops a chip-integrated diamond photonic crystal cavity embedding an ensemble of NV centers. The diamond nanostructure is integrated with a SiN optical waveguide that is connected to an optical fiber. We demonstrate cryogenic operation through the observation of Purcell-enhanced photon emission using the edge-coupled fiber. The operation of the integrated chip at low temperatures is a key step toward scalable quantum communication nodes based on diamond color centers.

	%%%%%%%%%%%%%%%%%%%%
	%\section{Device fabrication and integration}
	\begin{figure}
		\centering
		\includegraphics[width=85 mm]{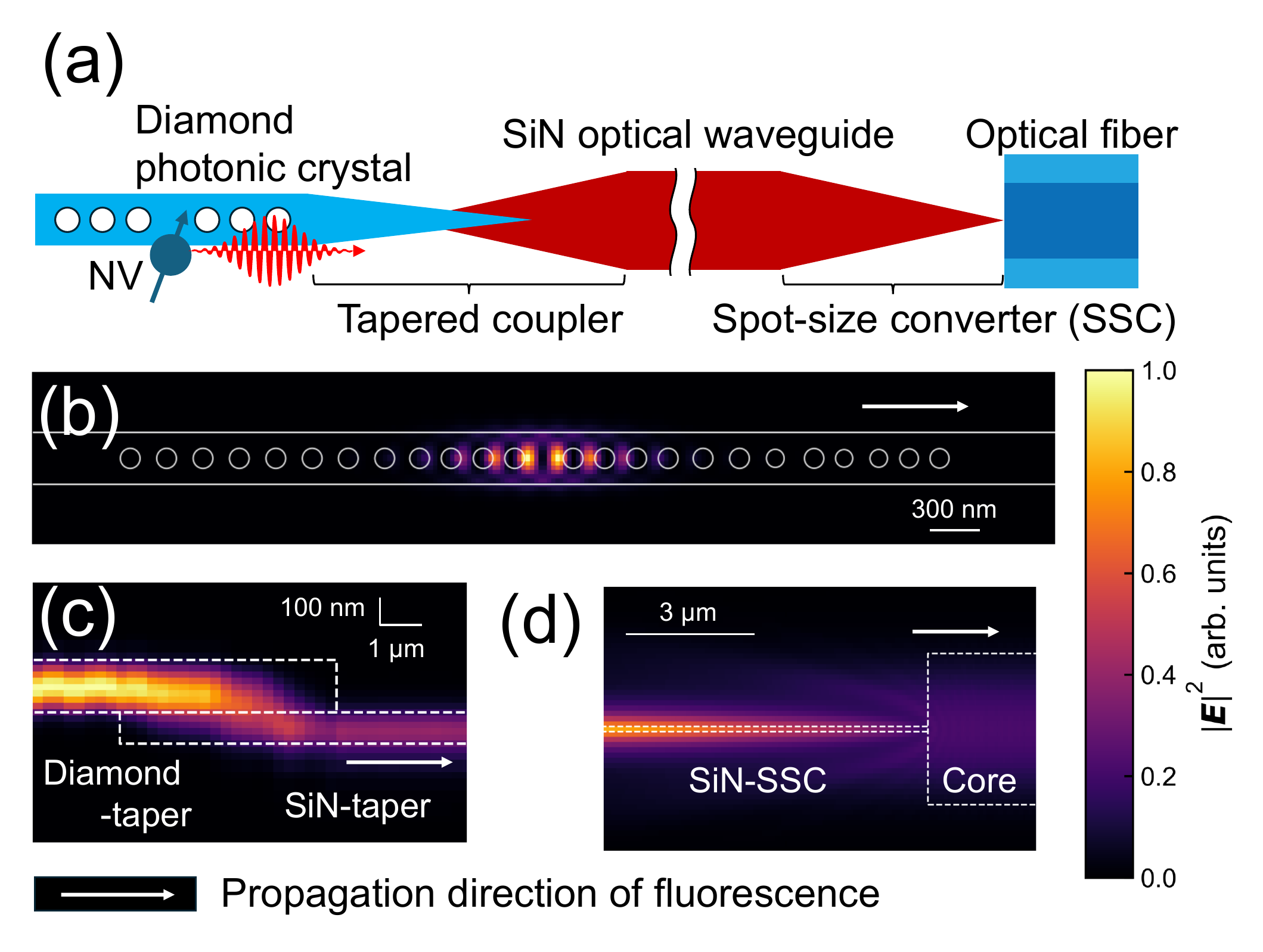}
		\caption
		{Schematic of the device. (a) Conceptual illustration of a chip-integrated diamond nanophotonic cavity embedding NV centers. The NV-cavity interaction enhances photon emission from the NV center into the SiN optical waveguide. The optical waveguide couples to an optical fiber at the chip edges via a spot-size converter (SSC). FDTD simulation of electric-field intensity of (b) the photonic crystal cavity (top view), (c) the diamond-SiN tapered coupler (side view), and (d) the SiN-fiber interface region (side view).
		} 
		\label{fig:schematic}
	\end{figure}

	The concept of the chip-integrated diamond device is illustrated in Figure~\ref{fig:schematic}(a). A diamond photonic crystal cavity (Figure~\ref{fig:schematic}(b)) embedding NV centers is placed on a tapered SiN optical waveguide. A fraction of the fluorescence from NV centers near the cavity center is directed toward the tapered region at one end of the cavity and is then adiabatically coupled to the tapered SiN waveguide (Figure~\ref{fig:schematic}(c)). The other end of the waveguide is coupled to an optical fiber via a spot-size converter (SSC) \cite{Hiraki2013} (Figure~\ref{fig:schematic}(d)). This enables the collection of the emitted photons through the fiber. %Although not relevant to this study, adjacent electrodes can apply both DC and microwave fields to control the spin or orbital state.

	Next, we describe the design of the photonic crystal cavity (width: 300 nm; thickness: 200 nm). The electric-field intensity distribution simulated using the finite difference time domain (FDTD) method, at the transversal electric (TE) resonance mode is shown in Figure~\ref{fig:schematic}(b). Although not visible on the color scale, a small leakage of the electric field to the right-hand side occurs because of the modulation of the hole positions and sizes to enhance the light-extraction efficiency. The hole apodization is optimized such that the quality factor is reduced to approximately half of that of the original structure, corresponding to the critical-coupling condition (see Supporting Information S2 for details of the design). For the fundamental mode, quality factor $Q$ and mode volume $V_\mathrm{mode}$ are $(Q,\ V_\mathrm{mode}/(\lambda/n)^3)=(2.7\times10^4,\ 0.52)$ for the structure without hole optimization, $(1.7\times10^4,\ 0.52)$ for the structure with optimized holes, and $(1.0\times10^4,\ 0.53)$ for the structure with optimized holes and an SiO$_2$ layer underneath, respectively. The apodization reduces $Q$ by approximately 40$\%$ as intended, and the SiO$_2$ layer further decreases $Q$ by approximately 40$\%$. By contrast, $V_\mathrm{mode}$ changes only marginally.
	
\begin{figure}
	\centering
	\includegraphics[width=80 mm]{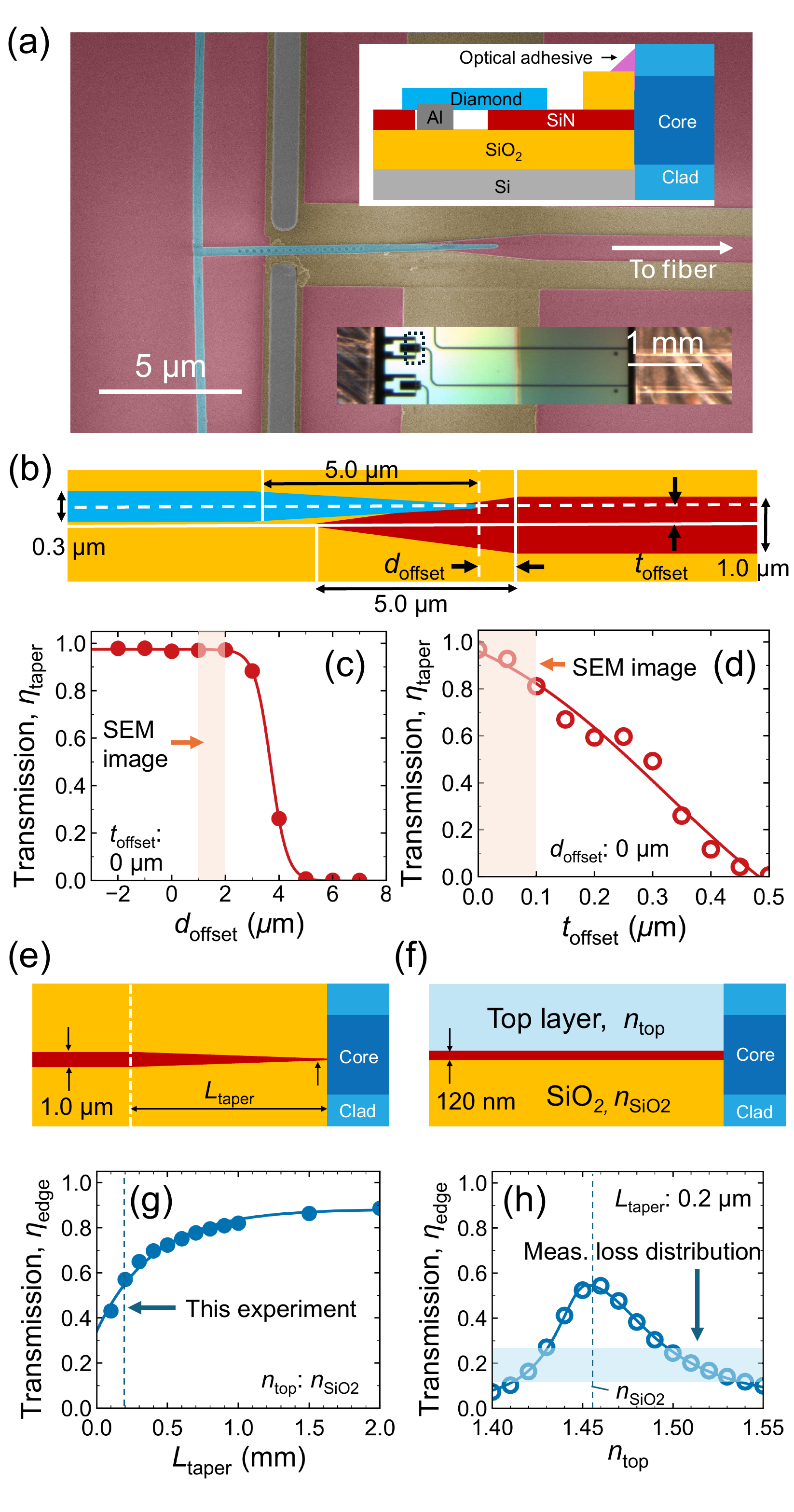}
	\caption
	{Photonic crystal integration with a SiN optical waveguide. (a) SEM image of an integrated diamond photonic crystal on an optical waveguide chip. The upper inset shows a side view of the device. The lower inset is a photograph of one optical waveguide, and the region enclosed by the dashed line contains the device shown in the SEM image. (b) Alignment tolerance of the optical transmission efficiency at the interface between the SiN waveguide and diamond tapers for (c) horizontal and (d) vertical offsets. The fit curves are based on a phenomenological hyperbolic-tangent function. The alignment accuracy estimated from the SEM image is indicated by the shaded regions. (e) Top view and (f) side view of the SiN-fiber edge coupler. Transmission efficiency at the SiN-fiber interface as a function of (g) the SiN taper length (fit: exponential function) and (h) the refractive index of the top material, $n_\mathrm{top}$ (fit: asymmetric Lorentzian function).	} 
	\label{fig:device}
	\end{figure}

	The device fabrication process comprises the following: (1) diamond nanofabrication, (2) fabrication of the SiN optical waveguide chip, and (3) integration of the diamond photonic crystal and optical waveguide using a pick-and-place technique. First, T-shaped diamond photonic crystals are fabricated from bulk diamond (type Ib; Sumitomo Electric) using a combination of anisotropic and quasi-isotropic etching (see Supporting Information S3 for details of the fabrication process) \cite{Mouradian2018}. To create NV centers in the photonic crystal, the sample is irradiated with an electron beam (2 MeV, 10 mA, $1 \times 10^{18}\,\mathrm{cm^{-2}}$), followed by thermal annealing at $1000\,^{\circ}\mathrm{C}$ for 2 h.  Meanwhile, a 1-$\mu$m-wide optical waveguide is fabricated from 120-nm-thick SiN on an SiO$_2$ (2 $\mu$m)/Si substrate at a standard silicon photonics foundry. To enhance optical transmission between the diamond and SiN waveguides, a 5-$\mu$m-long taper with a 100 nm tip width is formed at the ends of both the diamond and SiN waveguides. The SiN waveguide chip is mounted on a copper fixture, followed by the integration of a 24-channel single-mode fiber-block array using an optical adhesive. Before fiber integration, a 2--3-$\mu$m SiO$_2$ layer is deposited on the SiN waveguide to reduce loss due to the refractive-index mismatch between SiO$_2$ and the adhesive. Then, using a thin tungsten probe tip in a custom-built microscope, the diamond photonic crystal is picked and placed onto the optical waveguide. A scanning electron microscopy (SEM) image of the device, a schematic side view, and an optical microscope image of the waveguide, are shown in Figure~\ref{fig:device}(a).

	To evaluate the chip-integrated device embedding an ensemble of NV centers, we use the following experimental setup. For photoluminescence (PL) measurements, a 515-nm green laser is focused through an objective lens to off-resonantly excite the NV centers. PL is collected either with a free-space confocal setup or fiber-based setup. In both cases, the signal is transmitted to a spectrometer. A dichroic mirror and bandpass filter are used to suppress unwanted light from the 515-nm excitation laser. For relaxation-time measurements, an avalanche photodiode (APD), instead of the spectrometer, detects photons, and a time tagger records the photon arrival times from the APD (see Supporting Information S1 for details of the experimental setup). %Cryogenic measurements are performed using a dilution refrigerator.

	After chip integration, we first evaluate the transmission loss through the optical waveguide, which is an important figure of merit for integrated quantum photonic chips sensitive to photon loss events. For the proposed device, losses arise from the diamond-SiN interface, propagation loss in the SiN waveguide, and the SiN-fiber interface. First, the transmission efficiency at the diamond-SiN taper interface, $\eta_\mathrm{taper}$, is estimated as follows. Based on the simulated alignment tolerance (Figure~\ref{fig:device}(b)--(d)), the vertical offset, $t_\mathrm{offset}$, is the dominant contributor to the coupling loss. Based on the SEM images (Figure~\ref{fig:device}(a)), $t_\mathrm{offset}$ is estimated to be within 100 nm, which corresponds to a transmission efficiency above 80$\%$. 
	Next, the propagation loss of the SiN waveguide is estimated to be 1.9 dB/cm based on the measurements on waveguides with different lengths, leading to $85\%$ transmission for the proposed $\sim$0.35-cm-long waveguide. Finally, the transmission efficiency at the fiber edge coupler (the SiN-fiber interface), $\eta_\mathrm{edge}$, is $19.7\pm4.5\%$ (-7 dB), based on two-port transmission measurements. The uncertainty is the standard deviation over 14 samples; the maximum (minimum) value is 26.6$\%$ (10.9$\%$). The averaged performance is approximately one-fifth (-6.5 dB) of the simulated value (90$\%$) (Figure~\ref{fig:device}(e)--(g)). We attribute the loss to a short SSC length ($\sim$60$\%$ (-2 dB) based on the simulation; Figure~\ref{fig:device}(g)), a refractive-index mismatch between SiO$_2$ and the adhesive ($\sim$60$\%$ (-2 dB) based on simulation and experiments; Figure~\ref{fig:device}(h)), and misalignment between the fiber and waveguide ($\sim$70$\%$ (-1.5 dB) based on experiments). Although the origin of the additional 1-dB loss is uncertain, it may arise from an underestimation of the refractive-index mismatch of the adhesive and from loss at the fiber connectors.
	
	%%%%%%%%%%%%%%%%%%%%
	%\section{Results and discussions}
	
	\begin{figure}[t]
	\centering
	\includegraphics[width=85 mm]{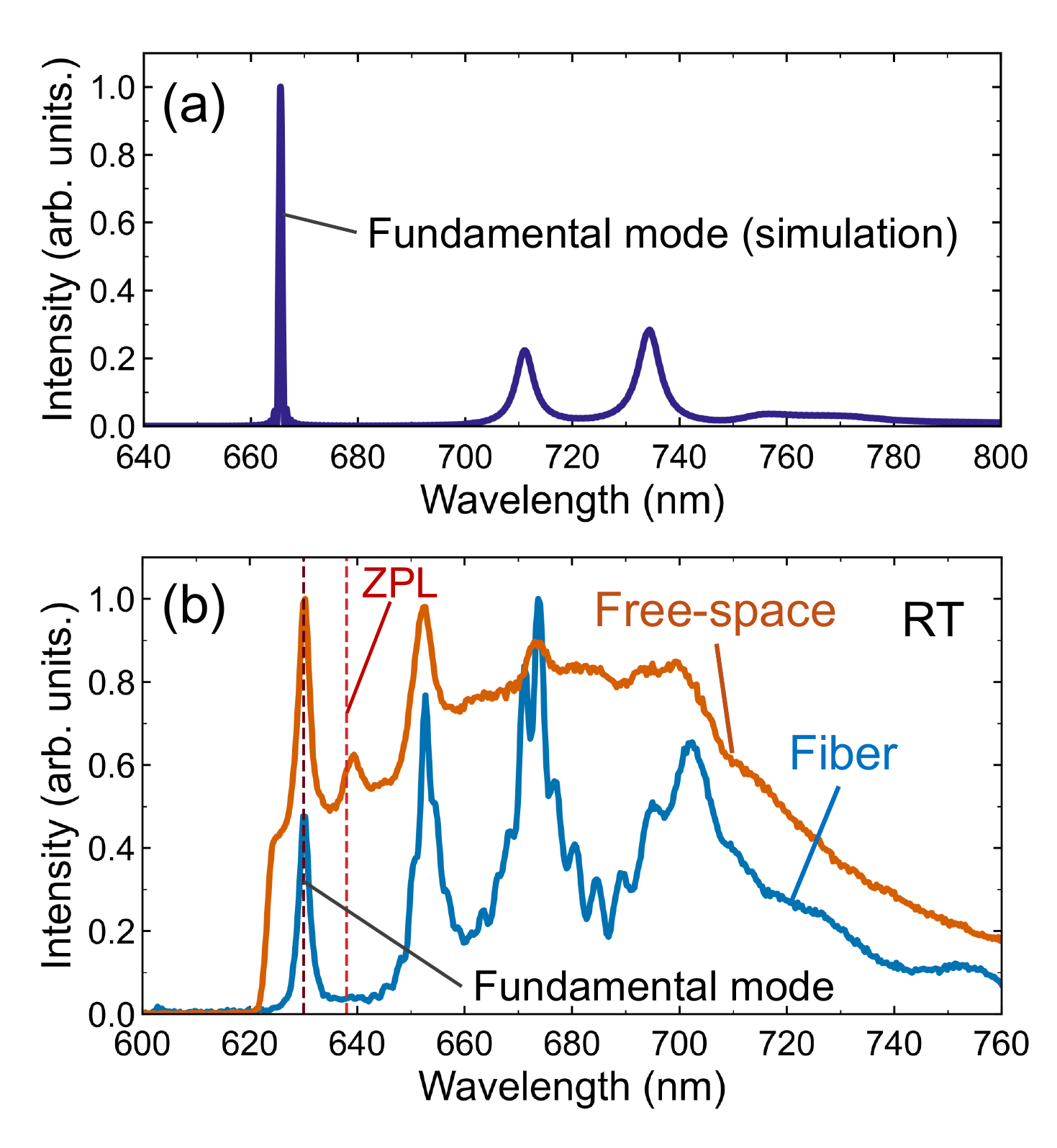}
	\caption
	{Photonic crystal cavity evaluation. (a) Simulated transmission spectrum for TE excitation applied outside the cavity and monitored inside the cavity. (b) Photoluminescence (PL) spectrum of NV centers at the center of the cavity. PL signals are collected via both a free-space setup and the integrated fiber at room temperature.		
	} 
	\label{fig:PL}
	\end{figure}
	
	Next, to characterize the photonic crystal cavity and transmission spectrum along the fiber collection path, we measure the PL spectrum of the NV centers. We compare the PL spectra of NV centers in the photonic crystal cavity obtained using the free-space setup, fiber-based setup, and simulation (Figure~\ref{fig:PL}). For both setups, several resonance peaks are observed on top of the broad phonon sideband (PSB) emission spectrum spanning 620--800 nm. The zero-phonon line (ZPL) of the NV centers appears at 637 nm in the free-space measurement but is almost absent in the fiber-based measurement. This is because the fiber collection path predominantly detects the cavity-waveguide channel and thus emphasizes cavity-coupled emission rather than the total free-space emission, leading to a strong suppression of the direct (non-cavity-coupled) ZPL feature, in good agreement with the simulation (Figure~\ref{fig:PL}(a)). Notably, in the simulation, we do not focus on the absolute resonance frequency, because we can finely tune the resonance frequency experimentally by varying the processing conditions. The leftmost peak in Figure~\ref{fig:PL}(b) is identified as the target cavity mode because it exhibits the highest $Q$ and matches the simulated spectrum. The quality factor before integration is $Q=420\pm110$, and it decreases to $Q=190\pm70$ after integration, where the averages are taken over eight samples. We attribute the reduction in $Q$ to the SiO$_2$ layer beneath the photonic crystal structure. The larger excess loss in the experiment ($\Delta(1/Q)_\mathrm{SiO2}\sim2.9\times10^{-3}$) than in the simulation ($\Delta(1/Q)_\mathrm{SiO2}\sim4.1\times10^{-5}$) may arise from finite material absorption, surface roughness, and fabrication-induced defects of the SiO$_2$ layer.

	\begin{figure}[t]
		\centering
		\includegraphics[width=80 mm]{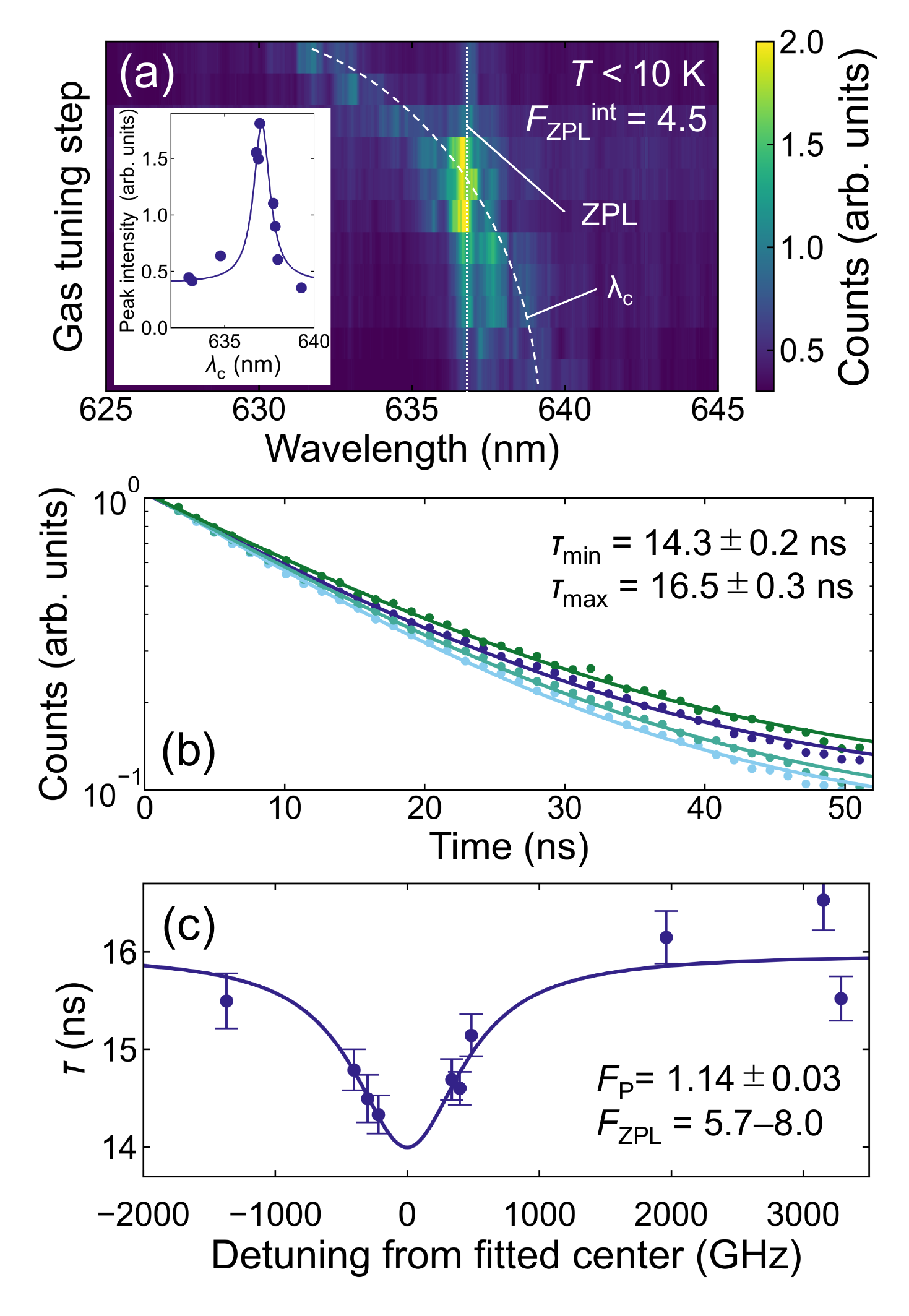}
		\caption
		{Purcell-enhanced fluorescence and relaxation dynamics. (a) PL spectra around the ZPL for different steps of gas tuning. An inset shows the peak intensity at ZPL as a function of the cavity wavelength, $\lambda_\mathrm{c}$, where the background is subtracted. The cavity center wavelength is extracted from a combination of Lorentzian and Gaussian fits. (b) Time-resolved PL traces at different detunings. (c) Relaxation time, $\tau$, as a function of detuning, $\Delta$.
		} 
		\label{fig:Purcell}
	\end{figure}
	%%%%%%%%%%%%%%%%%%%%

	Following the room-temperature measurements, we evaluate the interaction between the optical cavity and NV centers via the Purcell effect. Measurements are performed at cryogenic temperatures below 10 K (see Supporting Information S1 for details). A gas-tuning technique is used to tune the cavity into resonance with the NV-center ZPL. First, nitrogen gas is introduced into the dilution refrigerator. Adsorption of nitrogen onto the diamond photonic crystal increases the effective refractive index of the device, thereby increasing the resonant wavelength (i.e., decreasing the resonant frequency). After shifting the resonance to longer wavelengths, irradiation with the 515-nm green laser locally heats the device and induces nitrogen desorption. 	
	The resonant wavelength is shifted to shorter wavelengths by adjusting the laser power and irradiation time (Figure~\ref{fig:Purcell}(a)). When the cavity is tuned into resonance with the ZPL (637 nm), resonantly enhanced emission is observed. The Purcell factor from the enhancement of the ZPL intensity is expressed as $F_\mathrm{ZPL}^\mathrm{int}=I_\mathrm{ZPL}^\mathrm{on}/I_\mathrm{ZPL}^\mathrm{off}=4.5$, as shown in the inset of Figure~\ref{fig:Purcell}(a), where $I_\mathrm{ZPL}^\mathrm{on}$ is the maximum PL intensity on resonance and $I_\mathrm{ZPL}^\mathrm{off}$ is the PL intensity far off resonance \cite{Crook2020}.

	Additionally, we measure the relaxation time of the NV centers as a function of detuning (Figures~\ref{fig:Purcell}(b) and (c)) to evaluate the Purcell factor from a different perspective. As the NV--cavity detuning, $\Delta$, decreases, the relaxation time, $\tau$, also decreases (Figure~\ref{fig:Purcell}(c)), indicating the presence of NV--cavity interaction. Here, we measure only the ZPL emission using a bandpass filter. We fit $\tau(\Delta)$ using the following model \cite{Codreanu2025}:
	\begin{equation}\label{eq:tau_fit}
		\tau(\Delta) = \frac{\tau_1}{C f(\Delta)+1},
	\end{equation}
	where $\tau_1=1/\gamma_1$ is the NV lifetime in the far-detuned regime, $C=4g_0^2/(\kappa\gamma_1)$ is the cooperativity, $f(\Delta)=1/(1+4\Delta^2/\kappa^2)$ is the spectral-mismatch factor, $g_0$ is the optical vacuum coupling rate between the NV center and cavity, $\kappa$ is the total cavity decay rate, and $\gamma_1$ is the NV decay rate in the far-detuned regime. The Purcell factor is defined as $F_\mathrm{P}=\gamma_\mathrm{P}/\gamma_1=C+1$, where $\gamma_\mathrm{P}$ is the total radiative decay rate of the NV center on resonance. In the absence of NV-cavity interaction, $C=0$ and $F_\mathrm{P}=1$. The fit yields $C = 0.14\pm0.03$ ($F_\mathrm{P}=1.14$), $\kappa/2\pi=940\pm370$ GHz ($Q=510^{+320}_{-150}$), and $\tau_1=15.9\pm0.2$ ns. The fitted minimum relaxation time is $\tau_\mathrm{min}^\mathrm{fit}=14.0$ ns. %Meanwhile, the maximum (minimum) relaxation time obtained from individual relaxation traces is $\tau_\mathrm{max}=16.5\pm0.3$ ns ($\tau_\mathrm{min}=14.3\pm0.2$ ns).

	The deviation between $F_\mathrm{ZPL}^\mathrm{int}$ and $F_\mathrm{P}$ is due to the low Debye--Waller factor, $\eta_\mathrm{DW}$, of NV centers, 2--3\% \cite{Li2015c}. In this case, $F_\mathrm{P}$ and $C$ do not directly correspond to the enhancement of ZPL photon emission. 
	To explicitly account for the Debye--Waller factor, we define the ZPL cooperativity $C_\mathrm{ZPL}$ and ZPL Purcell factor $F_\mathrm{ZPL}$ as $F_\mathrm{ZPL}=C_\mathrm{ZPL}+1=\gamma_\mathrm{P,ZPL}/\gamma_\mathrm{ZPL}$, where $\gamma_\mathrm{P,ZPL}$ is the cavity-enhanced ZPL decay rate and $\gamma_\mathrm{ZPL}=\eta_\mathrm{DW}\gamma_1$ is the ZPL decay rate. Then, $C_\mathrm{ZPL}=4g_0^2/(\kappa\gamma_\mathrm{ZPL})=C\gamma_1/\gamma_\mathrm{ZPL}=C/\eta_\mathrm{DW}$. Thus, $C_\mathrm{ZPL}=4.7$--7.0 and $F_\mathrm{ZPL}=5.7$--8.0, depending on the Debye--Waller factor. These values correspond to the resonant enhancement of the ZPL emission rate, which agrees well with $F_\mathrm{ZPL}^\mathrm{int}$. Compared with state-of-the-art demonstrations using NV centers that reported $C_\mathrm{ZPL}\sim 60$ ($Q\sim 3300$) \cite{Li2015c}, $C_\mathrm{ZPL}\sim 5$ in the proposed device is mainly limited by the quality factor ($Q\sim 500$). This limitation can be mitigated by advances in fabrication techniques \cite{Codreanu2025} or using a diamond membrane \cite{Ding2024}. Additionally, the decrease in $Q$ after the pick-and-place process should be addressed by updating the design of the SiN waveguide chip.
	
Finally, we discuss the NV--cavity vacuum coupling rate, which characterizes the intrinsic interaction strength between the NV centers and cavity. Using the fitted parameters, we obtain $g_0^\mathrm{exp}/2\pi=0.57$ GHz. For comparison, assuming ideal dipole alignment with the cavity electric field at an antinode, we estimate $g_0^\mathrm{theory}/2\pi=(d_{\perp,\mathrm{zpl}}/2\pi)E_\mathrm{zpf}\simeq3.0$ GHz for a cavity with $V_\mathrm{mode}/(\lambda/n)^3=0.5$. Here, the zero-point fluctuation electric field is $E_\mathrm{zpf}=\sqrt{\hslash \omega_\mathrm{r}/(2\epsilon V_\mathrm{mode})}\simeq 5.8\times10^3$ V/cm ($\omega_\mathrm{r}/2\pi=475$ THz and $\epsilon=5.7\epsilon_0$ for diamond). The ZPL transition dipole moment is taken as $d_{\perp,\mathrm{ZPL}}=\sqrt{\eta_\mathrm{DW}}\,d_\perp$, where $d_\perp$ is the total transition dipole moment (including both ZPL and PSB). We use $d_{\perp,\mathrm{ZPL}}/2\pi\simeq0.5$ MHz/(Vcm$^{-1}$), obtained from the spontaneous-emission relation $\gamma_1=\omega_\mathrm{NV}^3 (d_\perp^{[\mathrm{C\cdot m}]})^2/(3\pi\epsilon_0\hslash c^3)$, where $\omega_\mathrm{NV}$ is the optical transition frequency of the NV center, $d_\perp^{[\mathrm{C\cdot m}]}/\hslash=d_\perp$, $\hslash$ is the Dirac's constant, $\epsilon_0$ is the vacuum permittivity, and $c$ is the speed of light.

The deviation between $g_0^\mathrm{exp}$ and $g_0^\mathrm{theory}$ can be explained by two main factors. First, dipole-orientation mismatch with the cavity field and spatial/orientational averaging over an NV ensemble reduce the effective coupling. Assuming an isotropic and uniform dipole distribution within the cavity mode volume, $g_0^\mathrm{theory}$ is reduced by a factor of $\sim$0.3--0.4 (see Supporting Information S7 for detailed calculations). Second, uncertainty in cavity linewidth $\kappa$ affects the extracted $g_0^\mathrm{exp}$. Because gas tuning reduces $Q$, and $Q$ near resonance cannot easily be determined because of spectral overlap with the ZPL and etalon fringes in the fiber-detection path, $Q$ may be underestimated. In a worst-case scenario where the actual $Q$ is $\sim250$ (half of 500), $g_0^\mathrm{exp}$ would be underestimated by a factor of $\sqrt{2}$. Therefore, we attribute the observed deviation in $g_0$ mainly to ensemble averaging and uncertainty in $\kappa$.

	In conclusion, we have developed a fiber-integrated diamond photonic crystal cavity embedding NV centers. Using the device, we have  demonstrated cryogenic operation in a dilution refrigerator by correcting photoluminescence through the optical fiber and observing Purcell enhancement, thereby confirming NV--cavity coupling in a fully fiber-integrated cryogenic device platform. The current transmission efficiency from the diamond taper to the edge-coupled fiber is $\sim10\%$, and we expect that it can be improved to $\sim$40$\%$ by optimizing the SSC length and refractive index of the optical adhesive. The ZPL Purcell factor is estimated to be 5--8, primarily limited by the optical quality factor.  Further improvements in diamond nanofabrication and integration should increase $Q$ and enable high ZPL enhancement, advancing scalable quantum photonic technologies based on color centers in diamond, such as single-photon sources, quantum repeaters, and microwave-to-optical transducers.

	%%%%%%%%%%%%%%%%%%%%
	\section*{acknowledgment}
	H. Kosaka acknowledges the funding support from the Japan Science and Technology Agency (JST) Moonshot R$\&$D grant (JPMJMS2062) and a JST CREST grant (JPMJCR1773). H. Kosaka also acknowledges the Ministry of Internal Affairs and Communications (MIC) for funding the research and development to construct a global quantum cryptography network  (JPMI00316), and the Japan Society for the Promotion of Science (JSPS) Grants-in-Aid for Scientific Research (20H05661, 20K20441). This work is also supported by Japan Science and Technology Agency (JST) as part of Adopting Sustainable Partnerships for Innovative Research Ecosystem (ASPIRE) (JPMJAP24C1).
	
	%%%%%%%%%%%%%%%%%%%%	
	%\section*{Data availability}
	%The data underlying this study are openly available in [Repository Name] at [Persistent Link to data in Repository, e.g., DOI, Accession Number].
	%%%%%%%%%%%%%%%%%%%%
\bibliographystyle{achemso}
%\bibliography{C:/Users/scott/OneDrive/document/kosakaLab/RefRenamed/library}
\providecommand{\latin}[1]{#1}
\makeatletter
\providecommand{\doi}
  {\begingroup\let\do\@makeother\dospecials
  \catcode`\{=1 \catcode`\}=2 \doi@aux}
\providecommand{\doi@aux}[1]{\endgroup\texttt{#1}}
\makeatother
\providecommand*\mcitethebibliography{\thebibliography}
\csname @ifundefined\endcsname{endmcitethebibliography}
  {\let\endmcitethebibliography\endthebibliography}{}

%\includepdf[pages=-]{C:/Users/scott/OneDrive/document/kosakaLab/paper/Chip_integrated_diamond_cc/tex/chip_integrated_diamond_supply_01/chip_integrated_diamond_supply_01.pdf}

\end{document}

\title{Supplementary Materials for Scalable on-chip integration of diamond color centers for cryogenic quantum photonics}
	
	\author{H. Kurokawa}
\email{kurokawa-hodaka-hm@ynu.ac.jp}
\affiliation{Quantum Information Research Center, Institute of Advanced Sciences, Yokohama National University, 79-5 Tokiwadai, Hodogaya, Yokohama, Kanagawa, 240-8501, Japan}

\author{K.~Sato}
\affiliation{Graduate School of Engineering Science, Yokohama National University, 79-5 Tokiwadai, Hodogaya, Yokohama, Kanagawa, 240-8501, Japan}

\author{M.~Kamata}
\affiliation{Graduate School of Engineering Science, Yokohama National University, 79-5 Tokiwadai, Hodogaya, Yokohama, Kanagawa, 240-8501, Japan}

%	\author{T.~Kobayashi}
%\affiliation{Graduate School of Engineering Science, Yokohama National University, 79-5 Tokiwadai, Hodogaya, Yokohama, Kanagawa, 240-8501, Japan}

%	\author{K.~Goto}
%\affiliation{Graduate School of Engineering Science, Yokohama National University, 79-5 Tokiwadai, Hodogaya, Yokohama, Kanagawa, 240-8501, Japan}

%	\author{M.~Yamamoto}
%\affiliation{Graduate School of Engineering Science, Yokohama National University, 79-5 Tokiwadai, Hodogaya, Yokohama, Kanagawa, 240-8501, Japan}

\author{S.~Ishida}
\affiliation{Research Center for Advanced Science and Technology (RCAST), The University of Tokyo, 4-6-1 Komaba, Meguro-ku, Tokyo, 153-8904, Japan}

\author{H.~Matsukiyo}
\affiliation{Institute of Industrial Science (IIS), The University of Tokyo, 4-6-1 Komaba, Meguro-ku, Tokyo, 153-8904, Japan}

\author{N.~Pholsen}
\affiliation{Institute of Industrial Science (IIS), The University of Tokyo, 4-6-1 Komaba, Meguro-ku, Tokyo, 153-8904, Japan}

\author{M.~Nishioka}
\affiliation{Institute of Industrial Science (IIS), The University of Tokyo, 4-6-1 Komaba, Meguro-ku, Tokyo, 153-8904, Japan}

\author{S.~Ji}
\affiliation{Institute of Industrial Science (IIS), The University of Tokyo, 4-6-1 Komaba, Meguro-ku, Tokyo, 153-8904, Japan}
\affiliation{Research Institute for Semiconductor Engineering, Hiroshima University, 1-4-2 Kagamiyama, Higashi-Hiroshima, Hiroshima, 739-8527, Japan}

\author{H.~Otsuki}
\affiliation{Institute of Industrial Science (IIS), The University of Tokyo, 4-6-1 Komaba, Meguro-ku, Tokyo, 153-8904, Japan}

\author{S.~Hachuda}
\affiliation{Graduate School of Engineering Science, Yokohama National University, 79-5 Tokiwadai, Hodogaya, Yokohama, Kanagawa, 240-8501, Japan}

\author{M.~Kunii}
\affiliation{Graduate School of Engineering Science, Yokohama National University, 79-5 Tokiwadai, Hodogaya, Yokohama, Kanagawa, 240-8501, Japan}

\author{T.~Tamanuki}
\affiliation{Graduate School of Engineering Science, Yokohama National University, 79-5 Tokiwadai, Hodogaya, Yokohama, Kanagawa, 240-8501, Japan}

%\author{T.~Makino}
%\affiliation{Quantum Information Research Center, Institute of Advanced Sciences, Yokohama National University, 79-5 Tokiwadai, Hodogaya, Yokohama, Kanagawa, 240-8501, Japan}
%\affiliation{Advanced Power Electronics Research Center, AIST, Onogawa, Tsukuba, Ibaraki, 305-8569, Japan}

%\author{M.~Diego}
%	\affiliation{Institute of Industrial Science (IIS), The University of Tokyo, 4-6-1 Komaba, Meguro-ku, Tokyo, 153-8904, Japan}

%\author{H.~Oka}
%\affiliation{Institute of Industrial Science (IIS), The University of Tokyo, 4-6-1 Komaba, Meguro-ku, Tokyo, 153-8904, Japan}

\author{K.~Kimura}
\affiliation{Quantum Materials and Applications Research Center (QUARC), QST, 1233 Watanuki-machi, Takasaki, Gunma, 370-1292, Japan}

\author{K.~Takenaka}
\affiliation{Quantum Materials and Applications Research Center (QUARC), QST, 1233 Watanuki-machi, Takasaki, Gunma, 370-1292, Japan}

\author{Y.~Sekiguchi}
\affiliation{Quantum Information Research Center, Institute of Advanced Sciences, Yokohama National University, 79-5 Tokiwadai, Hodogaya, Yokohama, Kanagawa, 240-8501, Japan}

%\author{H.~Kato}
%\affiliation{Quantum Information Research Center, Institute of Advanced Sciences, Yokohama National University, 79-5 Tokiwadai, Hodogaya, Yokohama, Kanagawa, 240-8501, Japan}
%\affiliation{Advanced Power Electronics Research Center, AIST, Onogawa, Tsukuba, Ibaraki, 305-8569, Japan}

%\author{K.~Koshino}
%\affiliation{Quantum Information Research Center, Institute of Advanced Sciences, Yokohama National University, 79-5 Tokiwadai, Hodogaya, Yokohama, Kanagawa, 240-8501, Japan}
%\affiliation{Institute for Liberal Arts, Institute of Science Tokyo, 2-8-30 Kokonodai, Ichikawa, Chiba 272-0827, Japan}

\author{S.~Onoda}
\affiliation{Quantum Information Research Center, Institute of Advanced Sciences, Yokohama National University, 79-5 Tokiwadai, Hodogaya, Yokohama, Kanagawa, 240-8501, Japan}
\affiliation{Quantum Materials and Applications Research Center (QUARC), QST, 1233 Watanuki-machi, Takasaki, Gunma, 370-1292, Japan}

%\author{M.~Nomura}
%\affiliation{Quantum Information Research Center, Institute of Advanced Sciences, Yokohama National University, 79-5 Tokiwadai, Hodogaya, Yokohama, Kanagawa, 240-8501, Japan}
%\affiliation{Institute of Industrial Science (IIS), The University of Tokyo, 4-6-1 Komaba, Meguro-ku, Tokyo, 153-8904, Japan}

\author{S.~Iwamoto}
\affiliation{Quantum Information Research Center, Institute of Advanced Sciences, Yokohama National University, 79-5 Tokiwadai, Hodogaya, Yokohama, Kanagawa, 240-8501, Japan}
\affiliation{Research Center for Advanced Science and Technology (RCAST), The University of Tokyo, 4-6-1 Komaba, Meguro-ku, Tokyo, 153-8904, Japan}
\affiliation{Institute of Industrial Science (IIS), The University of Tokyo, 4-6-1 Komaba, Meguro-ku, Tokyo, 153-8904, Japan}

\author{T.~Baba}
\affiliation{Quantum Information Research Center, Institute of Advanced Sciences, Yokohama National University, 79-5 Tokiwadai, Hodogaya, Yokohama, Kanagawa, 240-8501, Japan}
\affiliation{Graduate School of Engineering Science, Yokohama National University, 79-5 Tokiwadai, Hodogaya, Yokohama, Kanagawa, 240-8501, Japan}

\author{H.~Kosaka}
\email{kosaka-hideo-yp@ynu.ac.jp}
\affiliation{Quantum Information Research Center, Institute of Advanced Sciences, Yokohama National University, 79-5 Tokiwadai, Hodogaya, Yokohama, Kanagawa, 240-8501, Japan}
\affiliation{Graduate School of Engineering Science, Yokohama National University, 79-5 Tokiwadai, Hodogaya, Yokohama, Kanagawa, 240-8501, Japan}

	\maketitle
	
	\onecolumngrid
	
	\tableofcontents
	
	\beginsupplement
	%%%%%%%%%%%%%%%%%%%%
	\clearpage
	\section{Experimental setup}
	\subsection{Room temperature measurement}
\begin{figure}[b]
		\centering
		\includegraphics[width=100 mm]{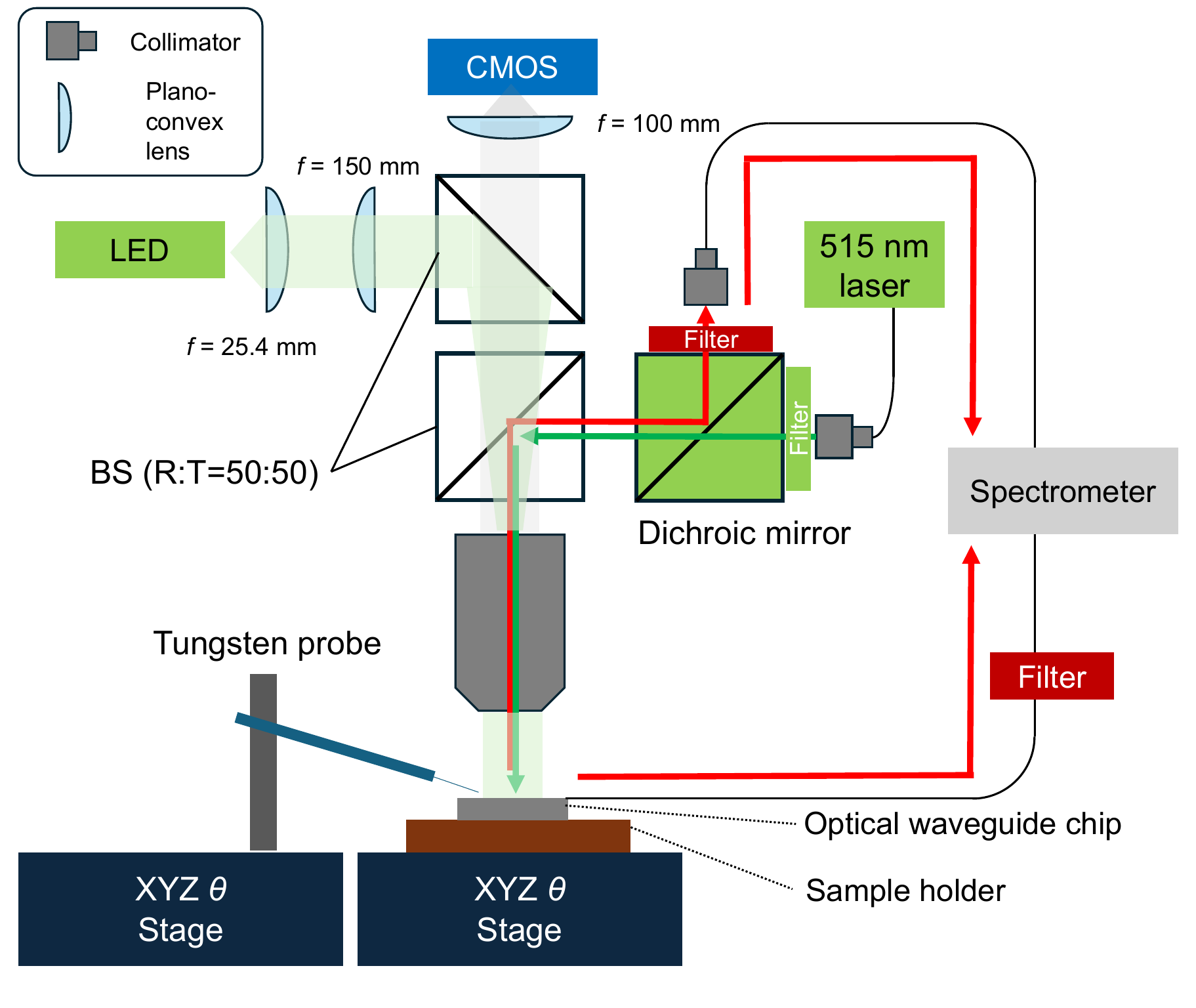}
		\caption
		{Schematic of the room-temperature experimental setup for pick-and-place integration and photoluminescence measurements. %An LED provides coaxial illumination to monitor the sample with a CMOS camera. A thin tungsten probe and four-axis  stages enable sample positioning and pick-and-place transfer of photonic crystals. A 515-nm laser is used to off-resonantly excite the NV centers. The PL is collected through both a free-space path and a fiber-based path and sent to a spectrometer. 
		} 
		\label{fig:RT_setup}
	\end{figure}

	Figure~\ref{fig:RT_setup} shows a schematic of the room-temperature pick-and-place device integration and photoluminescence (PL) measurement setup. The sample is illuminated by an LED (M530L4, Thorlabs) through an objective lens (RMS4X/RMS20X, Thorlabs/LMPLFLN 100X, NA=0.8, Olympus) and monitored with a CMOS camera (CS165MU/M, Thorlabs). A thin tungsten probe (TP-0002/TP-0005+TPH-10, MicroSupport) is manipulated using four-axis (XYZ+$\theta$) stages (NanoMax MAX313D/M + CR1 (RP01/M), Thorlabs), enabling pick-up of a photonic crystal and placement onto an optical waveguide. For PL measurements, a 515-nm laser (06-MLD, Cobolt) (spectral clean-up filter: FF02-520/25-5$\times$5, Semrock) is focused onto the sample through an objective lens (LMPLFLN 100X, NA=0.8, Olympus) to nonresonantly excite the NV centers. The PL is collected through both a free-space path (dichroic mirror: FF552-Di02-5$\times$8, bandpass filter: FF01-709/167-5$\times$5, Semrock) and a fiber-based path (longpass filter: FF01-593/LP-25, Semrock) and detected with a spectrometer (CCT12, Thorlabs/QEpro, Ocean Optics). Unwanted light from the 515-nm laser is suppressed using an appropriate combination of a dichroic mirror and a longpass filter.

	\subsection{Dilution refrigerator measurement}
\begin{figure}
		\centering
		\includegraphics[width=120 mm]{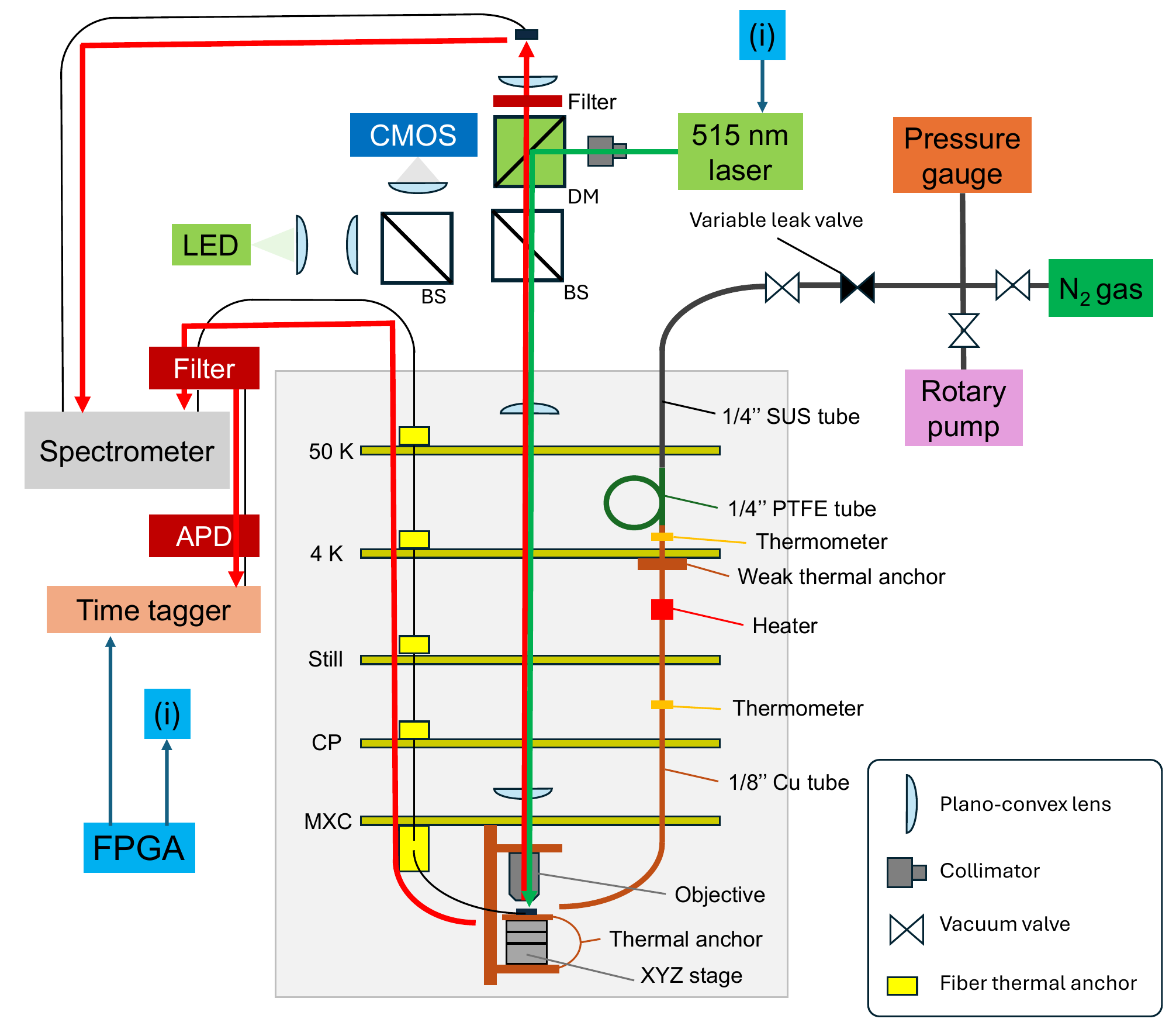}
		\caption
		{Schematic of an experimental setup for low temperature measurements using a dilution refrigerator. 
		} 
		\label{fig:dilution_setup}
	\end{figure}

	Figure~\ref{fig:dilution_setup} shows a schematic of the experimental setup for cryogenic measurements in a dilution refrigerator (LD400, Bluefors). The setup consists of (i) sample imaging and positioning, (ii) off-resonant optical excitation, (iii) PL collection through free-space and fiber paths, and (iv) spectral and time-resolved detection.

	For sample imaging and alignment, an LED and a CMOS camera are used to illuminate and monitor the sample position. The sample is positioned using a three-axis piezo stage (ANPx101/RES/LT + ANPz101/RES/LT, Attocube) to align the device with the 515-nm excitation laser (06-MLD, Cobolt) spot. A cryogenic-compatible objective (LT-APO/VIS/0.82, NA = 0.82, Attocube) is used for both excitation and free-space collection.

	The PL is collected through both a free-space path (longpass filter: FELH0550, Thorlabs) and a fiber-based path (longpass filter for wide range PL measurements: FF01-593/LP-25, Semrock/ bandpass filter for ZPL measurements: FF01-637/7-25, Semrock) and sent to room-temperature detection optics. Inside the dilution refrigerator, six single-mode fibers (visible $\times4$: S630-HP, Thorlabs; C-band $\times2$: SMF28, Thorlabs) are installed and thermally anchored at each temperature stage. In the measurements presented in the main text, we use an SMF28 (C-band) fiber for PL collection because the connector loss is lowest for this fiber in our system. Although SMF28 is designed for telecom wavelengths and is not strictly single-mode at visible wavelengths, the fiber length inside the cryostat is short and the measurement observables in this work (spectral features and relaxation dynamics) are not limited by modal purity.

	For PL spectrum measurements, we use a spectrometer (SpectraPro-2500i + PIXIS-100BRX-LFS, Princeton Instruments). For relaxation measurements, PL is detected with an avalanche photodetector (SPCM-AQRH-14-FC, Excelitas Technologies) and time-tagged using a time tagger (MultiHarp 150, PicoQuant) with a time-bin of 1.28 ns. An FPGA (PXIe-7820R, National Instruments) gates the 515-nm laser and provides triggers for the time tagger. The 515-nm laser fall time (80--20\%) is approximately 1.0 ns in our setup, which is negligible for the lifetime analysis in the main text.

	The nitrogen gas line used for cavity-frequency tuning is configured as follows. Nitrogen gas is supplied from a cylinder outside the dilution refrigerator. The pressure in the vacuum line connected to the refrigerator is monitored and controlled using a vacuum gauge (GP-1G/WP-16, ULVAC) and a rotary pump (G-10DA, ULVAC), and a variable leak valve (DN KF16 ISO-KF, Leybold) is used to finely control the nitrogen flow. Inside the dilution refrigerator, the gas line consists of a 1/4-inch SUS tube, a 1/4-inch PTFE tube, and a 1/8-inch copper tube connected by standard fittings (Swagelok; PTFE or brass). Two thermometers (W-LAK-PT-103+/WCX-1050-CU-HT-1.4L, Lakeshore) and a 50-$\Omega$ heater (W-LAK-HTR-50, Lakeshore) are attached to the copper tube to control its temperature. The PTFE section provides thermal isolation, while the copper tube is weakly anchored to the 4-K plate to balance heating controllability and base-temperature performance. The lowest temperature of the copper tube is approximately 10 K without heating. In this setup, the base temperature in the circulation mode of the dilution refrigerator is approximately 60 mK.

	For cavity-frequency tuning, we use the following procedure. First, the copper tube is heated using the heater (typically 0.2 A). After the thermometer above the 4-K plate exceeds 60 K and the thermometer between the still and CP plates exceeds 200 K, nitrogen gas is introduced by opening the leak valve (typical pressure $\sim$2000 Pa upstream of the valve). The MXC plate temperature increase to $\sim$ 10 K from 4 K due to the heating. After confirming a resonance shift due to nitrogen adsorption, we finely tune the cavity frequency by adjusting the 515-nm laser power and irradiation time. The typical tuning laser power at the sample is 1--10 mW.
	
	\section{Simulation}
	
	%\textcolor{red}{simulation condition}
	For the simulation shown in the main text, we use a finite difference time domain (FDTD) simulation software (Lumerical, Ansys). In addition to the simulation mentioned in the main text, the SiN optical waveguide width (1-$\mu$m) and thickness (120 nm) are optimized so that the transmission efficiency from SiO$_2$-cladded region to air-cladded region is maximized (95.4$\%$). Additionally, the SiN spot-size converter tip shape (103-nm width, 120-nm thickness) is optimized to maximize the transmission efficiency from a fiber to the spot-size converter (95.6$\%$). These simulation are performed at 637 nm. For the simulation including the cavity structure, the wavelength around resonance, 665--666 nm, is used for the simulation.

\begin{figure}
		\centering
		\includegraphics[width=120 mm]{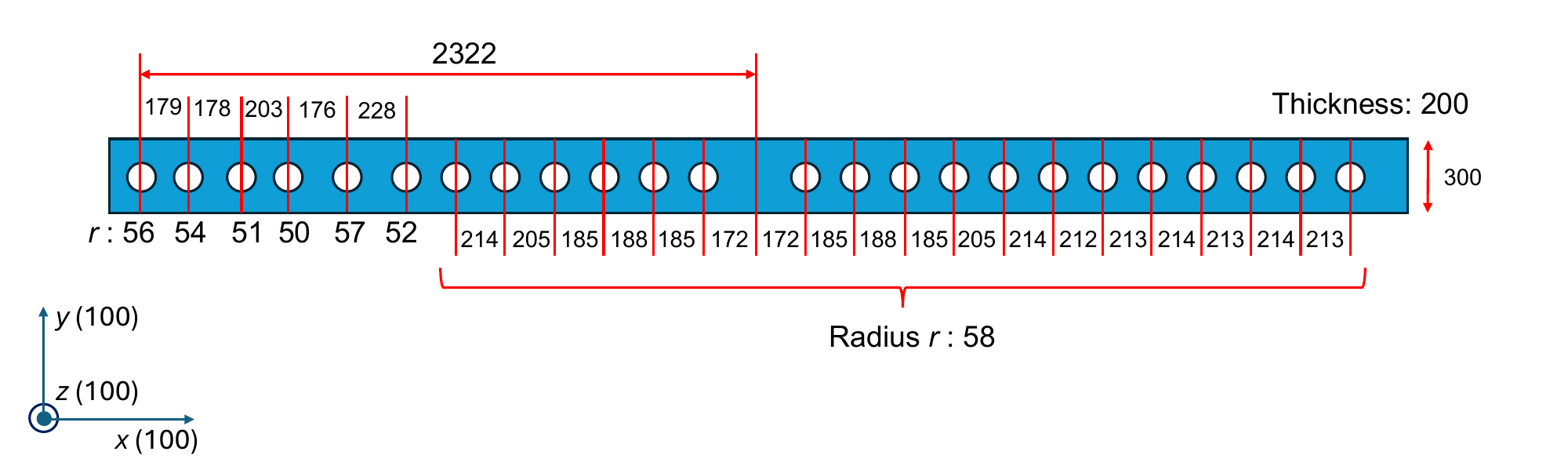}
		\caption
		{Optimized hole size and positions of the photonic crystal used for the simulation in the main text. All units are in nanometer. The crystal direction and definition of the cartesian coordinates are also shown.
		} 
		\label{fig:PhC_design}
	\end{figure}

	Figure \ref{fig:PhC_design} shows the design of the diamond photonic crystal cavity with optimized hole sizes and positions. There are twenty-four circular holes in total. Hole radius and position are shown in Fig. \ref{fig:PhC_design}. At the center of the photonic crystal, there is a defect region to confine electric field. The six holes on the left are apodized to enhance the light extraction efficiency from the center of the cavity. 
	
	\section{Diamond photonic crystal fabrication}
\begin{figure}
		\centering
		\includegraphics[width=120 mm]{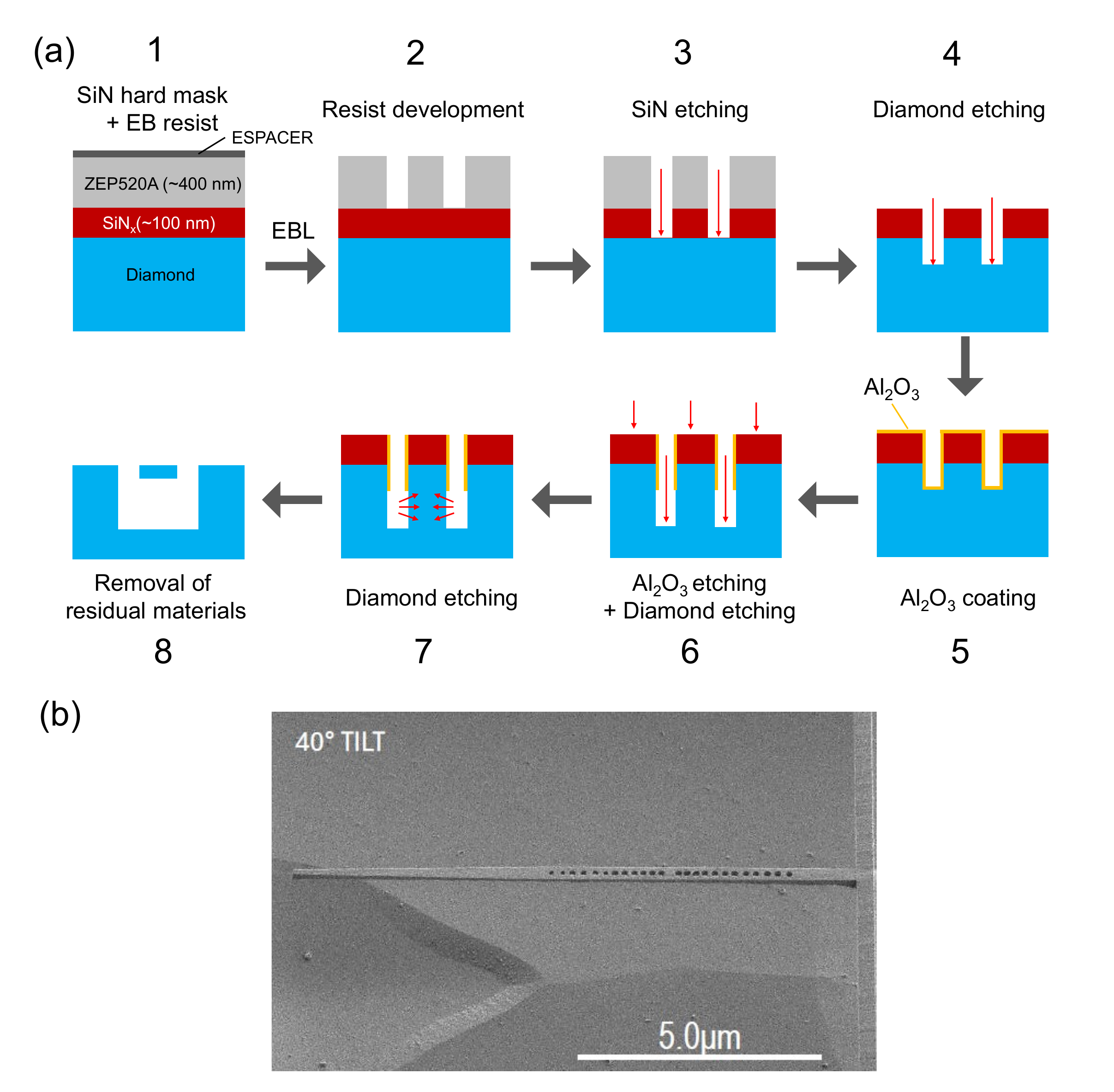}
		\caption
		{Fabrication of diamond photonic crystal cavities. (a) Fabrication process. (b) An SEM image of a fabricated device.
		} 
		\label{fig:PhC_fab}
	\end{figure}

	%\textcolor{red}{Information will be provided from Iwamoto-sensei}
In this section, we describe the fabrication process of the diamond photonic crystal cavity [Fig.~\ref{fig:PhC_fab}(a)]. First, a 100-nm-thick SiN$_x$ hard mask is deposited on a diamond substrate by chemical vapor deposition (300 $^\circ$C, 90 Pa, 100 W). The substrate is then coated with ZEP520A resist ($\sim$400 nm) and baked at 180 $^\circ$C for 40 min, followed by a coating of ESPACER 300Z and baking at 110 $^\circ$C for 10 min. Next, the photonic crystal pattern is defined by electron-beam lithography (EBL) at an acceleration voltage of 100 keV. The resist is developed using pure water and ZED-N50, followed by a rinse in ZND-B. The pattern is transferred to the SiN layer by inductively coupled plasma reactive ion etching (ICP-RIE) with CHF$_3$:O$_2$ = 9:1 (ICP/BIAS = 200/50 W, 1.5 Pa, 60 s). The diamond substrate is then etched by ICP-RIE in O$_2$ (ICP/BIAS = 300/30 W, 0.5 Pa, 300 s). To protect the cavity structure during the subsequent isotropic etching step, an approximately 20-nm-thick Al$_2$O$_3$ layer is deposited by atomic layer deposition at 200 $^\circ$C. The Al$_2$O$_3$ layer is subsequently etched by ICP-RIE with CF$_4$:Ar = 5.5:6.0 (ICP/BIAS = 125/500 W, 0.75 Pa, 40 s), after which the diamond is etched again by O$_2$ ICP-RIE (ICP/BIAS = 300/30 W, 0.5 Pa, 300 s). Air-suspended structures are then formed by isotropic etching in O$_2$ (ICP/BIAS = 900/0 W, 2 Pa, $\sim$2 h). Finally, the remaining SiN and Al$_2$O$_3$ are removed using vapor hydrofluoric acid (HF) at 32 $^\circ$C for 1 h and 35 $^\circ$C for 0.5 h. An SEM image of a fabricated device is shown in Fig.~\ref{fig:PhC_fab}(b).
		
	\section{Fiber-integrated chip}
	
\begin{figure}
		\centering
		\includegraphics[width=120 mm]{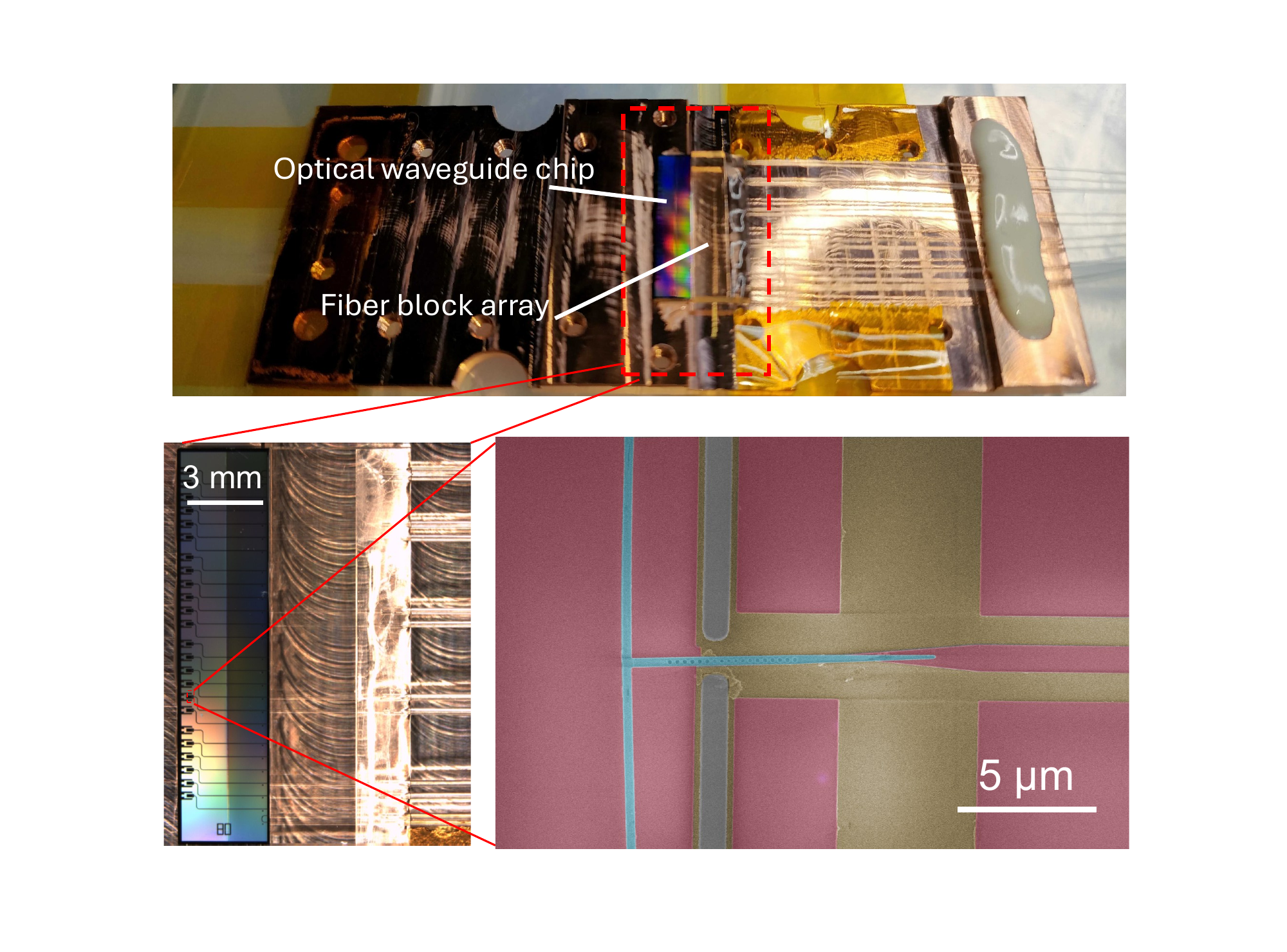}
		\caption
		{Images of an integrated device. The optical waveguide chip is mounted on a copper fixture, and the fiber-block array is fixed to the chip using an optical adhesive. The photonic crystal is placed on the SiN optical waveguide. 
		} 
		\label{fig:integrated_chip}
	\end{figure}

	Figure~\ref{fig:integrated_chip} provides an overview of the integrated chip. As explained in the main text, the optical waveguide chip is mounted on a copper fixture. The fiber-block array is attached to the chip using an optical adhesive. The optical waveguide chip can accommodate up to 24 channels (24 photonic crystals). In this experiment, however, we integrated only 12 fibers for cost reasons: four for alignment and eight for the measurements.
	
	\section{Cryogenic stability}
	
\begin{figure}
		\centering
		\includegraphics[width=80 mm]{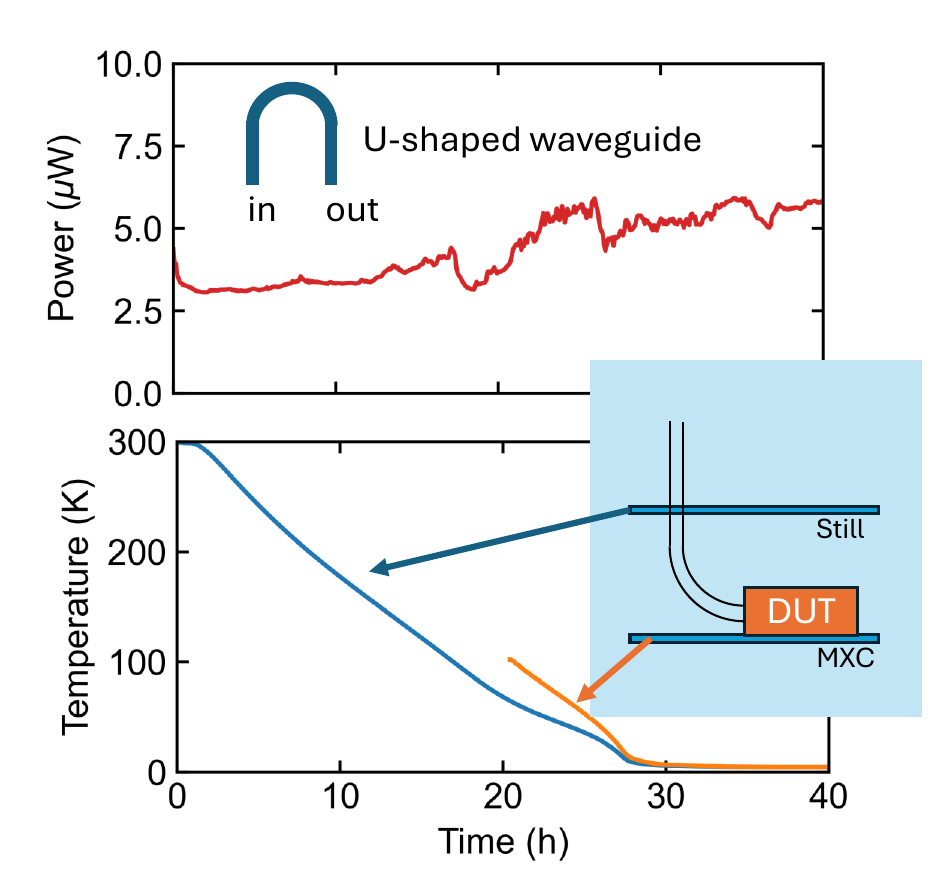}
		\caption
		{(Top) Transmission power as a function of elapsed time. (Bottom) Temperature as a function of elapsed time. Blue line is temperature at the still plate. Orange line is temperature at the mixing chamber plate.
		} 
		\label{fig:cryogenic_transmission}
	\end{figure}

	Figure \ref{fig:cryogenic_transmission} shows the transmitted power through an U-shaped test SiN waveguide integrated with fibers as a function of cooling time and temperature of each stage. The device is mounted onto the mixing chamber plate. Although there is a slight increase in the transmitted power, this device is stable down to 4 K. We use acrylate optical adhesive (AT6001, NTT-AT) to fix the fiber-block array with the optical waveguide chip \cite{Hiraki2013}. Although we test some  other optical adhesives, only the sample glued with AT6001 maintained stable transmission down to 4 K.
	
\section{The Purcell effect}
In this section, we derive the equation used to calculate the Purcell factor in the main text. We present both phenomenological and microscopic derivations.

\subsection{Phenomenological treatment}
\subsubsection{Definition}
The Purcell factor, $F_\mathrm{P}$, is defined as
\begin{equation}\label{eq:def_Fp}
	F_\mathrm{P}=\frac{\gamma_\mathrm{P}}{\gamma_1},
\end{equation}
where $\gamma_\mathrm{P}$ is the energy decay rate of an atom resonant with a cavity, and $\gamma_1$ is the decay rate of the atom in free space without a cavity (or under far-detuned conditions). Thus, $F_\mathrm{P}$ corresponds to the enhancement in the decay rate due to the interaction between an atom and a cavity. When there is no interaction between them, $F_\mathrm{P}=1$.

We also define the cooperativity, $C$, which is another widely used figure of merit for the atom-cavity interaction, as
\begin{equation}\label{eq:def_C_phenomenological}
	C = \frac{\gamma_\mathrm{cav}}{\gamma_1},
\end{equation}
where $\gamma_\mathrm{cav}$ corresponds to the decay rate of an atom via a coupled cavity. Then, assuming that the decay rate of the atom is governed by $\gamma_1$ and $\gamma_\mathrm{cav}$,
\begin{equation}\label{eq:def_gamma}
	\gamma_\mathrm{P} = \gamma_1 + \gamma_\mathrm{cav}.
\end{equation}
By combining Eqs. \eqref{eq:def_C_phenomenological} and \eqref{eq:def_gamma} with Eq. \eqref{eq:def_Fp},
\begin{equation}
	F_\mathrm{P} = 1 + C.
\end{equation}

\subsubsection{Experiment}
Experimentally, we usually observe the decay rate of an excited state, $\gamma_1^\mathrm{exp}=1/\tau_1^\mathrm{exp}$, through fluorescence decay. We model the decay channels of solid-state defects as
\begin{equation}
	\gamma_1^\mathrm{exp}=\gamma_\mathrm{rad}+\gamma_\mathrm{nonrad},
\end{equation}
where $\gamma_\mathrm{rad}$ is the radiative decay rate, and $\gamma_\mathrm{nonrad}$ is the nonradiative decay rate, including intersystem crossing. Furthermore,
\begin{equation}
	\gamma_\mathrm{rad}=\gamma_\mathrm{ZPL}+\gamma_\mathrm{PSB},
\end{equation}
where $\gamma_\mathrm{ZPL}$ is the zero-phonon-line (ZPL) decay rate, and $\gamma_\mathrm{PSB}$ is the phonon-sideband (PSB) decay rate. By defining the quantum efficiency, $\eta_\mathrm{QE}$, as
\begin{equation}
	\eta_\mathrm{QE} = \frac{\gamma_\mathrm{rad}}{\gamma_\mathrm{rad}+\gamma_\mathrm{nonrad}},
\end{equation}
and the Debye-Waller factor, $\eta_\mathrm{DW}$, as
\begin{equation}
	\eta_\mathrm{DW}=\frac{\gamma_\mathrm{ZPL}}{\gamma_\mathrm{ZPL}+\gamma_\mathrm{PSB}},
\end{equation}
$\gamma_1^\mathrm{exp}$ can be expressed as
\begin{equation}\label{eq:gamma_exp_1}
	\begin{aligned}
		\gamma_1^\mathrm{exp}&=
		\gamma_\mathrm{rad} 
		+\frac{1-\eta_\mathrm{QE}}{\eta_\mathrm{QE}}\gamma_\mathrm{rad}
		\\
		&=\frac{1}{\eta_\mathrm{QE}}\gamma_\mathrm{rad}
		\\
		&=\frac{1}{\eta_\mathrm{QE}\eta_\mathrm{DW}}\gamma_\mathrm{ZPL}.
	\end{aligned}
\end{equation}
Then, we assume that the interaction with a cavity enhances only $\gamma_\mathrm{ZPL}$ by a factor of $F_\mathrm{ZPL}$ when the cavity and the atom are on resonance, thereby modifying Eq. \eqref{eq:gamma_exp_1} as
\begin{equation}\label{eq:gamma_on_res}
	\gamma_1^\mathrm{exp,on} = F_\mathrm{ZPL}\gamma_\mathrm{ZPL}+\gamma_\mathrm{PSB}+\gamma_\mathrm{nonrad},
\end{equation}
\begin{equation}\label{eq:gamma_off_res}
	\gamma_1^\mathrm{exp,off} = \gamma_\mathrm{ZPL}+\gamma_\mathrm{PSB}+\gamma_\mathrm{nonrad}.
\end{equation}
By taking the difference between Eqs. \eqref{eq:gamma_on_res} and \eqref{eq:gamma_off_res},
\begin{equation}
	\gamma_1^\mathrm{exp,on}
	-
	\gamma_1^\mathrm{exp,off}
	= 
	(F_\mathrm{ZPL}-1)\gamma_\mathrm{ZPL}.
\end{equation}
Hence, by defining $C_\mathrm{ZPL} = F_\mathrm{ZPL}-1$,
\begin{equation}\label{eq:C_zpl_before_simplify}
	\begin{aligned}
		C_\mathrm{ZPL}
		&=\frac{1}{\gamma_\mathrm{1}^\mathrm{exp}\eta_\mathrm{QE}\eta_\mathrm{DW}}\qty(\gamma_1^\mathrm{exp,on}-\gamma_1^\mathrm{exp,off})
		\\
		&
		=\frac{\tau_\mathrm{1}^\mathrm{exp}}{\eta_\mathrm{QE}\eta_\mathrm{DW}}\qty(\frac{1}{\tau_1^\mathrm{exp,on}}-\frac{1}{\tau_1^\mathrm{exp,off}})
		.
	\end{aligned}
\end{equation}
Here, we use $\gamma_\mathrm{ZPL}=\gamma_\mathrm{1}^\mathrm{exp}\eta_\mathrm{QE}\eta_\mathrm{DW}$. Ideally, $\gamma_1^\mathrm{exp}=\gamma_1^\mathrm{exp,off}$ by definition. Thus,
\begin{equation}\label{eq:C_zpl_simplify}
	\begin{aligned}
		C_\mathrm{ZPL}
		&=\frac{1}{\eta_\mathrm{QE}\eta_\mathrm{DW}}\qty(\gamma_1^\mathrm{exp,on}/\gamma_1^\mathrm{exp,off}-1)
		\\
		&=
		\frac{1}{\eta_\mathrm{QE}\eta_\mathrm{DW}}\qty(\frac{\tau_1^\mathrm{exp,off}}{\tau_1^\mathrm{exp,on}}-1).
	\end{aligned}
\end{equation}
Eqs. \eqref{eq:C_zpl_before_simplify} and \eqref{eq:C_zpl_simplify} are commonly used to determine $C_\mathrm{ZPL}$ experimentally. Additionally, assuming $\tau_1^\mathrm{exp,off}=1/\gamma_1$ and $\tau_1^\mathrm{exp,on}=1/\gamma_\mathrm{P}$,
\begin{equation}
	C_\mathrm{ZPL}
	=
	\frac{1}{\eta_\mathrm{QE}\eta_\mathrm{DW}}
	\qty(F_\mathrm{P}-1)=\frac{C}{\eta_\mathrm{QE}\eta_\mathrm{DW}}.
\end{equation}
In the main text and the following analysis, we approximate $\eta_\mathrm{QE}\simeq1$ \cite{Ruf2021}.

\subsubsection{General case}
Consider the case in which incorporating an atom into a cavity modulates the decay rates as $\gamma_\mathrm{ZPL}\rightarrow\gamma_\mathrm{ZPL}'$, $\gamma_\mathrm{PSB}\rightarrow\gamma_\mathrm{PSB}'$, and $\gamma_\mathrm{nonrad}\rightarrow\gamma_\mathrm{nonrad}'$, due to the existence of a photonic bandgap or a nanocrystalline structure. These are the decay rates under far-detuned conditions. Then, based on the assumption that the atom-cavity interaction affects only $\gamma_\mathrm{ZPL}'$ on resonance,
\begin{equation}\label{eq:gamma_on_res_gen}
	\gamma_1^\mathrm{exp,on} = F_\mathrm{ZPL}\gamma_\mathrm{ZPL}'+\gamma_\mathrm{PSB}'+\gamma_\mathrm{nonrad}',
\end{equation}
\begin{equation}\label{eq:gamma_off_res_gen}
	\gamma_1^\mathrm{exp,off} = \gamma_\mathrm{ZPL}'+\gamma_\mathrm{PSB}'+\gamma_\mathrm{nonrad}'.
\end{equation}
Hence,
\begin{equation}
	C_\mathrm{ZPL} = \frac{1}{\gamma_\mathrm{ZPL}'}
	\qty(
	\gamma_1^\mathrm{exp,on}-\gamma_1^\mathrm{exp,off}
	).
\end{equation}
Assuming that the Debye-Waller factor and the quantum efficiency are unchanged in the cavity under far-off-resonant conditions (in other words, the decay rates are suppressed or enhanced by the same coefficient, $\alpha$, as $\gamma_\mathrm{ZPL}=\alpha\gamma_\mathrm{ZPL}'$, $\gamma_\mathrm{PSB}=\alpha\gamma_\mathrm{PSB}'$,
$\gamma_\mathrm{nonrad}=\alpha\gamma_\mathrm{nonrad}'$),
\begin{equation}
	\gamma_\mathrm{ZPL}'=\eta_\mathrm{QE}\eta_\mathrm{DW}\gamma_\mathrm{1}^\mathrm{exp,off}.
\end{equation}
Thus, the same equation as Eq. \eqref{eq:C_zpl_simplify} is obtained:
\begin{equation}\label{eq:C_zpl_general}
	\begin{aligned}
		C_\mathrm{ZPL} &= \frac{\tau_1^\mathrm{exp,off}}{\eta_\mathrm{QE}\eta_\mathrm{DW}}
		\qty(\frac{1}{\tau_1^\mathrm{exp,on}}-\frac{1}{\tau_1^\mathrm{exp,off}}
		)
		\\
		&=
		\frac{1}{\eta_\mathrm{QE}\eta_\mathrm{DW}}
		\qty(
		\frac{\tau_1^\mathrm{exp,off}}{\tau_1^\mathrm{exp,on}}-1
		)
		.
	\end{aligned}
\end{equation}
It should be noted that some assumptions here are too strong to be generalized:
\begin{enumerate}
	\item only $\gamma_\mathrm{ZPL}'$ is subjected to the Purcell enhancement.
	\item the Debye-Waller factor and the quantum efficiency are unchanged in the cavity at a far-detuned condition ($\gamma_\mathrm{ZPL}=\alpha\gamma_\mathrm{ZPL}'$, $\gamma_\mathrm{PSB}=\alpha\gamma_\mathrm{PSB}'$,
	$\gamma_\mathrm{nonrad}=\alpha\gamma_\mathrm{nonrad}'$).
\end{enumerate}
Generally, the existence of a cavity can modulate each decay rate in a different manner; thus, we cannot use a single $\alpha$ to express changes in the decay rates. In addition, it is possible that a cavity hosts several resonance modes at different wavelengths. However, since it is usually difficult to estimate all of the decay rates from a single experiment, we typically use Eq. \eqref{eq:C_zpl_general} to estimate $C_\mathrm{ZPL}$.

\subsection{Microscopic treatment}
In this section, we derive the equation used to fit $\tau(\Delta)$ in the main text from the atom-cavity Hamiltonian. In addition, we discuss the relationship between the above phenomenological treatment and this microscopic treatment.

The atom-cavity interaction Hamiltonian can be written as
\begin{equation}
	\hat{H}/\hslash 
	= \omega_\mathrm{a}\hat{\sigma}^+\hat{\sigma}
	+\omega_\mathrm{c}\hat{c}^\dagger\hat{c}
	+g_0(\hat{\sigma}^+\hat{c}+\hat{\sigma}\hat{c}^\dagger)
	,
\end{equation}
where $\hat{\sigma}\ (\hat{c})$ is the atomic lowering (resonator annihilation) operator, and $g_0$ is the transverse vacuum coupling between an atom and the resonator. In the frame rotating at the resonator frequency,
\begin{equation}
	\hat{H}^\mathrm{rot}/\hslash 
	= -\Delta_\mathrm{a}\hat{\sigma}^+\hat{\sigma}
	+g_0(\hat{\sigma}^+\hat{c}+\hat{\sigma}\hat{c}^\dagger)
	%+\sqrt{\kappa_\mathrm{ext}}C
	%(\hat{c}+\hat{c}^\dagger)
	,
\end{equation}
where $\Delta_\mathrm{a}=\omega_\mathrm{c}-\omega_\mathrm{a}$. Decoherence is incorporated through Lindblad superoperators within the quantum-master-equation formalism as
\begin{equation}
	\dv{\rho}{t}
	= -\frac{i}{\hslash}[\hat{H}^\mathrm{rot},\rho] 
	+ \gamma_{1}\mathcal{D}_{\hat{\sigma}}(\rho)
	+2\gamma_\phi\mathcal{D}_{\hat{\sigma}^+\hat{\sigma}}(\rho)
	+
	\kappa
	\mathcal{D}_{\hat{c}}(\rho),
\end{equation}
\begin{equation}
	\mathcal{D}_{\hat{L}}(\rho)
	=\hat{L}\rho\hat{L}^\dagger
	-\frac{1}{2}\qty{\hat{L}^\dagger\hat{L},\rho},
\end{equation}
where $[\hat{A},\hat{B}]=\hat{A}\hat{B}-\hat{B}\hat{A}$ and $\qty\big{\hat{A},\hat{B}}=\hat{A}\hat{B}+\hat{B}\hat{A}$. The time evolution of an operator expectation value is obtained using
\begin{equation}
	\ev*{\hat{A}}=\mathrm{Tr}(\rho \hat{A}),\ 
	\dv{\ev*{\hat{A}}}{t}=\mathrm{Tr}\qty(\hat{A}\dv{\rho}{t}),	
\end{equation}
where $\ev{\cdot}$ denotes the expectation value of an operator. Then,
\begin{equation}\label{eq:dsigma_dt}
	\dv{\ev{\hat{\sigma}}}{t}
	= -\qty(-i\Delta_{\mathrm{a}}
	+\gamma_2)\,\langle \hat{\sigma}\rangle
	\;+\; i g_0\,\langle \hat{\sigma}_z \hat{c}\rangle,
\end{equation}	
\begin{equation}\label{eq:dc_dt}
	\dv{\ev{\hat{c}}}{t}
	= -\frac{\kappa}{2}\langle \hat{c}\rangle
	\;-\; i g_0\,\langle \hat{\sigma}\rangle
	,
\end{equation}	
where $\gamma_1=1/\tau_1$ [$\tau_1$ (= $T_1$): the relaxation time of the atom], $\gamma_2=1/T_2 = 1/2T_1 + \gamma_\phi$ ($T_2$: the coherence time of the atom; $\gamma_\phi$: the pure-dephasing rate of the atom), and $\kappa$ is the total loss rate of the resonator. 

Assuming $\ev{\hat{c}(t)}\propto \mathrm{exp}(-i\Delta_\mathrm{a}t)$ in Eq. \eqref{eq:dc_dt}, we obtain
\begin{equation}
	\ev{\hat{c}} = \frac{-ig_0}{\kappa/2-i\Delta_\mathrm{a}}
	\ev{\hat{\sigma}}
	,
\end{equation}
and assuming no correlation between $\hat{\sigma}_z$ and $\hat{c}$, such that $\ev{\hat{\sigma}_z\hat{c}}=\ev{\hat{\sigma}_z}\ev{\hat{c}}$, together with the weak-excitation condition $\ev{\hat{\sigma}_z}=-1$,
\begin{equation}
	\begin{aligned}
		\dv{\ev{\hat{\sigma}}}{t}
		&= -\qty(-i\Delta_{\mathrm{a}}
		+\gamma_2
		)\ev{\hat{\sigma}}
		-\frac{g_0^2}{\kappa/2-i\Delta_\mathrm{a}}
		\ev{\hat{\sigma}}
		\\
		&=
		-\qty(
		-i\Delta_{\mathrm{a}}
		+\gamma_2
		+\frac{g_0^2(\kappa/2+i\Delta_\mathrm{a})}{(\kappa/2)^2+\Delta_\mathrm{a}^2}
		)\ev{\hat{\sigma}}
		\\
		&=
		-\qty(-i\Delta_{\mathrm{a}}+
		\frac{i\Delta_\mathrm{a}g_0^2}{(\kappa/2)^2+\Delta_\mathrm{a}^2}
		+\gamma_2
		+\frac{g_0^2(\kappa/2)}{(\kappa/2)^2+\Delta_\mathrm{a}^2}
		)\ev{\hat{\sigma}}.
	\end{aligned}
\end{equation}
Thus, under the above assumptions, the population  decay-rate enhancement of the atom due to the cavity is
\begin{equation}
	\frac{g_0^2\kappa}{(\kappa/2)^2+\Delta_\mathrm{a}^2}.
\end{equation}
When the atom-cavity detuning is zero, $\Delta_\mathrm{a}=0$, this reduces to
\begin{equation}
	\frac{4g_0^2}{\kappa}=\gamma_\mathrm{cav}.
\end{equation}
Additionally, neglecting pure-dephasing and setting $\gamma_2=\gamma_1/2$, the relaxation time as a function of detuning is 
\begin{equation}\label{eq:microscopic_tau_detuning}
	\begin{aligned}
		\tau(\Delta_{\mathrm{a}}) &= \frac{1}{\gamma_1+	\frac{g_0^2\kappa}{(\kappa/2)^2+\Delta_\mathrm{a}^2}}
		\\
		&=
		\frac{\tau_1}{1+\frac{4g_0^2}{\kappa\gamma_1}\frac{1}{1+4\Delta_{\mathrm{a}}^2/\kappa^2}}
		\\
		&=
		\frac{\tau_1}
		{1+Cf(\Delta_\mathrm{a})}
		,
	\end{aligned}
\end{equation}
where 
\begin{equation}
	C =\frac{\gamma_\mathrm{cav}}{\gamma_1} =\frac{4g_0^2}{\kappa\gamma_1}.
\end{equation}
Then, we can define the total decay rate as
\begin{equation}
	\gamma_\mathrm{P} = \gamma_1 +\frac{4g_0^2}{\kappa},
\end{equation}
and the Purcell factor is
\begin{equation}
	\begin{aligned}
		F_\mathrm{P} &= \frac{\gamma_\mathrm{P}}{\gamma_1} 
		\\&= 1 + \frac{4g_0^2}{\kappa\gamma_1}
		\\&= 1 + C
		.
	\end{aligned}
\end{equation}
Additionally, from Eq. \eqref{eq:microscopic_tau_detuning}, $\tau(\infty)=\tau_1$ and $\tau(0)=\tau_1/(1+C)$. Thus,
\begin{equation}
	C = \frac{\tau(\infty)}{\tau(0)} - 1.
\end{equation}
This equation coincides with Eq. \eqref{eq:C_zpl_general} except for the Debye-Waller factor and the quantum efficiency.

	%%%%%%%%%%%%%%%%%%%%	
\section{The NV-cavity coupling rate}

In the main text, the experimentally obtained NV--cavity coupling rate, $g_0^\mathrm{exp}$, is lower than that expected in the ideal case. We attribute this reduction to the effects of the NV-center ensemble and to uncertainty in the quality-factor estimate. In this section, we provide a slightly more detailed explanation of the former point.

Theoretically \cite{Faraon2011,Kristensen2012,Reiserer2015b},
\begin{equation}\label{eq:g_0_general}
	g_0 = \bm{d}(\bm{r})\cdot\bm{E}(\bm{r}) =E_\mathrm{zpf}\ \bm{f}(\bm{r})\cdot\bm{d}(\bm{r}),
\end{equation}
\begin{equation}
	E_\mathrm{zpf}=\sqrt{\frac{\hslash\omega_\mathrm{c}}{2\epsilon(\bm{r}_\mathrm{m}) V_\mathrm{mode}}},
\end{equation}
\begin{equation}
	V_\mathrm{mode}=\frac{\int_V\mathrm{d}\bm{r}\epsilon(\bm{r}) |\bm{E}(\bm{r})|^2}{\max\limits_{\bm{r} \in V}[\epsilon(\bm{r})|\bm{E}(\bm{r})|^2]},
\end{equation}
\begin{equation}
	\bm{r}_\mathrm{m}=\mathrm{argmax}_{\bm{r} \in V}[\epsilon(\bm{r})|\bm{E}(\bm{r})|^2],\ |\bm{E}(\bm{r}_\mathrm{m})|= E_\mathrm{zpf}
\end{equation}
%\begin{equation}
%	\bm{f}(\bm{r})
%	=\frac{\bm{E}(\bm{r})}{|\bm{E}(\bm{r}_\mathrm{m})|},
%\end{equation}
\begin{equation}
	\bm{f}(\bm{r})
	=\frac{\bm{E}(\bm{r})}{E_\mathrm{zpf}},\  |\bm{f}(\bm{r})|\le1,
\end{equation}
where $\bm{E}(\bm{r})$ is the zero-point fluctuation  electric-field vector, normalized such that $\int_V \mathrm{d}\bm{r}\,\epsilon(\bm{r})|\bm{E}(\bm{r})|^2=\hslash\omega_\mathrm{c}/2$, $\bm{d}$ is the transition dipole moment vector expressed in units of $2\pi\times$Hz/(V m$^{-1}$), $E_\mathrm{zpf}$ is the zero-point electric-field amplitude, $\bm{r}_\mathrm{m}$ is the position at which the electric-field energy density is maximized, $\bm{f}(\bm{r})$ is the normalized cavity mode function, $\omega_\mathrm{c}$ is the cavity angular frequency, and $V_\mathrm{mode}$ is the cavity mode volume. Here, we restrict our discussion to cavities for which the position of maximum electric-field amplitude coincides with that of maximum electric-field energy density. It should also be noted that several normalization conventions for cavity field profiles exist \cite{Han2016}. For simplicity, we adopt the above definition in the numerical calculations below.

To simplify the analysis, we assume that a dipole is placed at $\bm{r}_\mathrm{m}$ and aligned with the local cavity field. Then, Eq. \eqref{eq:g_0_general} reduces to
\begin{equation}
	g_0 = d_\mathrm{\perp,ZPL}E_\mathrm{zpf},
\end{equation}
where $d_\mathrm{\perp,ZPL}/2\pi=|\bm{d}|/2\pi$ is the amplitude of the ZPL transition dipole moment expressed in units of Hz/(V m$^{-1}$). For our diamond photonic crystal cavity, using $V_\mathrm{mode}/(\lambda/n)^3\simeq0.5$, $\omega_\mathrm{c}/2\pi=475$ THz, and $\epsilon =5.7\epsilon_0$, we obtain $E_\mathrm{zpf}\sim5.8\times10^5$ V/m (5.8$\times10^3$ V/cm). Meanwhile, $d_\mathrm{\perp,ZPL}$ can be estimated from the equation for the spontaneous emission rate,
\begin{equation}
	\gamma_1=
	\frac{\omega_\mathrm{NV}^3 (d_{\perp}^{[\mathrm{C\cdot m}]})^2}{3\pi\epsilon_0\hslash c^3}
	,
\end{equation}
as
\begin{equation}
	d_{\perp}^{[\mathrm{C\cdot m}]}=\sqrt{\frac{3\pi\epsilon_0\hslash c^3\gamma_1}{\omega_\mathrm{NV}^3 }},
\end{equation}
where $\omega_\mathrm{NV}/2\pi=475$ THz is the optical transition frequency of the NV center, $\epsilon_0$ is the vacuum permittivity, and $c$ is the speed of light. Using $\gamma_1=1/\tau_1$ and $\tau_1=16$ ns for our device, we obtain $d_\perp^{[\mathrm{C\cdot m}]}=2.4\times10^{-29}$ C$\cdot$m (7.1 Debye). Here, we define $d_\perp^{[\mathrm{C\cdot m}]}/\hslash=d_\perp$.
Then, based on the definition of the ZPL decay rate, $\gamma_\mathrm{ZPL}=\eta_\mathrm{DW}\gamma_1$, the ZPL transition dipole moment, $d_\mathrm{\perp,ZPL}^{[\mathrm{C\cdot m}]}=\sqrt{\eta_\mathrm{DW}}d_\perp^{[\mathrm{C\cdot m}]}$, is 1--1.2 Debye ($d_\mathrm{\perp,ZPL}/2\pi=0.5$--$0.6$ MHz/(V cm$^{-1}$)). Using these values, we obtain
\begin{equation}
	g_0^\mathrm{theory}/2\pi = 2.9-3.5\ \mathrm{GHz}.
\end{equation}

\begin{figure}
	\centering
	\includegraphics[width=90 mm]{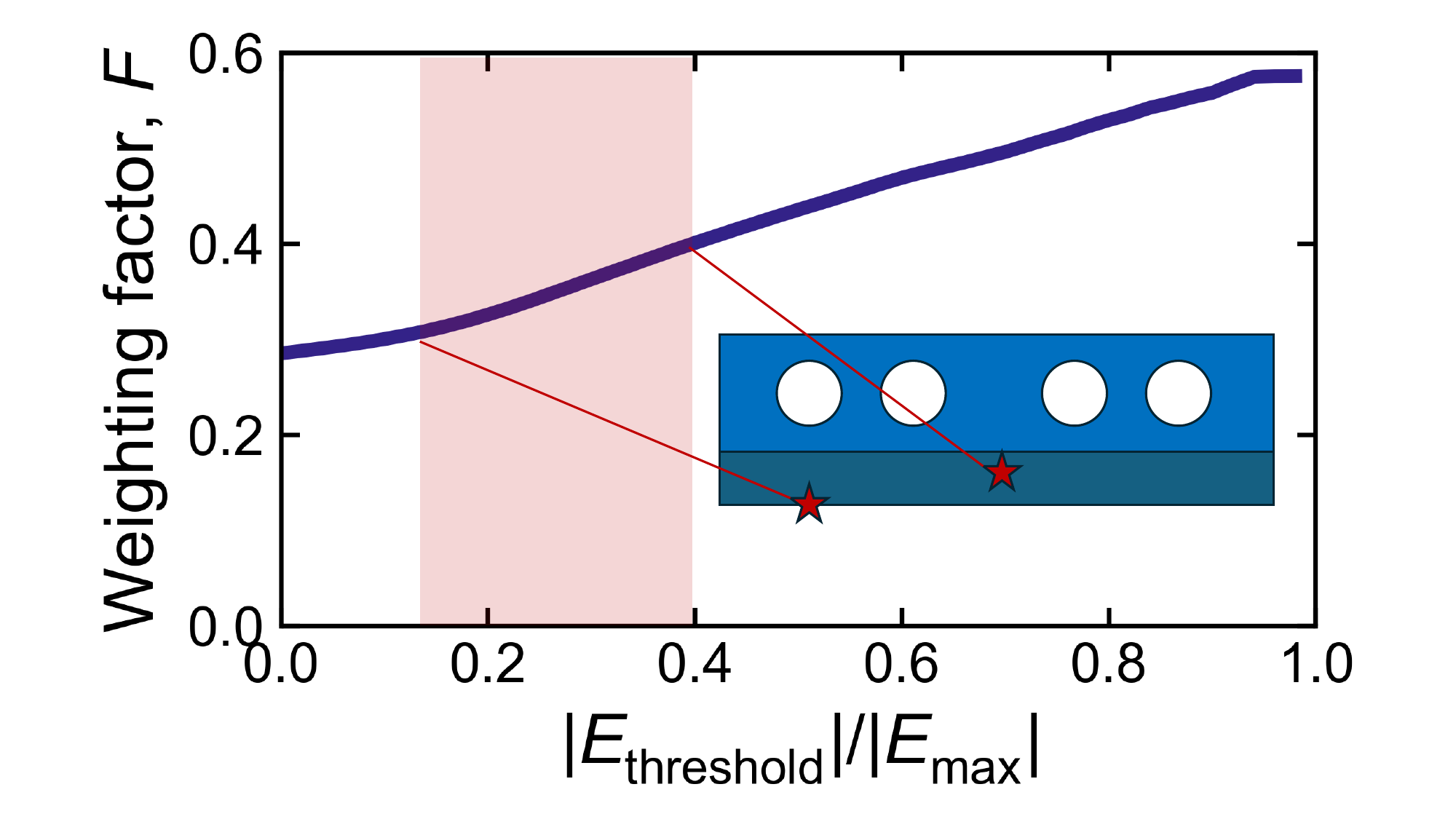}
	\caption
	{The weighting factor, $F$, in Eq.~\eqref{eq:total_weighting_factor} as a function of the threshold electric field normalized by the maximum electric field. The red shaded area indicates the possible range of $|E_\mathrm{threshold}|$. For example, when $\bm{r}=(0, 150, 0)$ nm and $(400, 150, -100)$ nm, corresponding to the star markers in the figure, $|E_\mathrm{threshold}|/|E_\mathrm{max}| \sim 0.4$ and $0.15$, respectively.
	} 
	\label{fig:weighting_factor}
\end{figure}

Compared with the experimentally estimated $g_0^\mathrm{exp}/2\pi=0.57$ GHz, this corresponds to a difference by a factor of approximately five to six. We attribute this deviation to the misalignment between the emission dipole orientation and the cavity electric field, as well as to uncertainty in the estimate of the $Q$ factor.
Since an ensemble of NV centers is distributed over the photonic crystal cavity region, we must consider the spatial electric-field distribution in the cavity. Assuming that the experimentally observed decay rate can be modeled  as a weighted average of the decay rates of individual emitters in the cavity, $\gamma_{1,i}$, (on resonance: $\gamma_{1,i}^\mathrm{on}$) as
\begin{equation}
	\gamma_1^\mathrm{exp,on} = \sum_i^N p_i \gamma_{1,i}^\mathrm{on},
\end{equation}
where $p_i$ ($\sum_i p_i=1$) is the weighting factor. Then, since
\begin{equation}
	1 + \frac{4(g_0^\mathrm{eff})^2}{\kappa\gamma_1}
	= 1 + \sum_i^N p_i \frac{4(g_{0,i})^2}{\kappa\gamma_1},
\end{equation}
we obtain
\begin{equation}
	g_0^\mathrm{eff}=\sqrt{\sum_i^N p_i E_\mathrm{zpf}^2 [\bm{f}(\bm{r}_i)\cdot\bm{d}(\bm{r}_i)]^2}.
\end{equation}
Assuming, for simplicity, that the emission dipole moments have equal magnitudes and are isotropically distributed,
\begin{equation}\label{eq:total_weighting_factor}
	g_0^\mathrm{eff}=
	g_0^\mathrm{theory}
	\sqrt{\sum_i^N p_i \frac{\qty[f_x(\bm{r}_i)^2+f_y(\bm{r}_i)^2+f_z(\bm{r}_i)^2]}{3}}
	=
	g_0^\mathrm{theory}\mathcal{F},
\end{equation}
where $\mathcal{F}$ is the spatially-averaged weighting factor. 
Additionally, we empirically assume that the contribution of each emitter is determined by the local electric-field amplitude. In a naive model, one would expect $p_i \propto |\bm{E}(\bm{r}_i)|/|E_\mathrm{max}|$, which would generally give a nonzero weight even for emitters located at the photonic crystal (PhC) edge because $|\bm{E}(\mathrm{edge})|>0$. However, our simulation suggests that the emitter weight empirically approaches zero near the PhC edge. To capture this behavior phenomenologically, we introduce a threshold electric field, $|E_\mathrm{threshold}|$, such that emitters at the PhC edge satisfy $w_i=0$, and hence $p_i=0$. We therefore define the unnormalized weight as
\begin{equation}
	w_i=
	\frac{\max\qty(|\bm{E}(\bm{r}_i)|-|E_\mathrm{threshold}|,\,0)}{|E_\mathrm{max}|},
\end{equation}
and the normalized weight as
\begin{equation}
	p_i=\frac{w_i}{\sum_j w_j}.
\end{equation}
Using these definitions, the total weighting factor, $F$, can be evaluated numerically by summing over the simulated electric-field distribution, as shown in Fig.~\ref{fig:weighting_factor}. The summation is performed over the region near the cavity center, $-400 < x < 400$ nm, $-150 < y < 150$ nm, and $-100 < z < 100$ nm. It is not straightforward to determine $|E_\mathrm{threshold}|$, because the electric-field amplitude at the PhC edges depends on how the edge positions are defined. When we set $|E_\mathrm{threshold}|/|E_\mathrm{max}| \sim 0.2$ based on the electric field at an edge $(\bm{r}=(300,150,0)\ \mathrm{nm})$, we obtain $F \simeq 0.3$. In that case, the effective $g_0$ can be reduced to approximately $0.3\,g_0^\mathrm{theory}$.

	%%%%%%%%%%%%%%%%%%%%
%\bibliography{C:/Users/scott/OneDrive/document/kosakaLab/RefRenamed/library,C:/Users/scott/OneDrive/document/kosakaLab/RefRenamed/book}
%Control: production of eprint (1) enabled
%